\documentclass{article}
\usepackage[a4paper,hmargin=1.2in,vmargin=1.2in]{geometry}
\usepackage[english]{babel}
\usepackage{graphicx}
\usepackage{amsfonts}
\usepackage{latexsym}
\usepackage{amsthm}
\usepackage{amssymb,amsmath}
\usepackage{framed}
\usepackage{color}
\usepackage{enumerate}
\usepackage{tikz}
\usepackage[hyphens]{url}

\usepackage{hyperref}

\newtheorem{Definition}{\sc Definition}[section]
\newtheorem{Theorem}[Definition]{\sc Theorem}
\newtheorem{Lemma}[Definition]{\sc Lemma}
\newtheorem{Proposition}[Definition]{\sc Proposition}
\newtheorem{Corollary}[Definition]{\sc Corollary}

\newtheorem{Remark}[Definition]{\sc Remark}

\newcommand{\BIGOP}[1]{\mathop{\mathchoice

{\raise-0.22em\hbox{\huge $#1$}}

{\raise-0.05em\hbox{\LARGE $#1$}}{\hbox{\large $#1$}}{#1}}}

\theoremstyle{remark}
\newtheorem{claim}{Claim}
\newtheorem{Proofc}{Proof of Claim}

\newenvironment{Proof}[1][]{\par \sc

Proof{#1}. \rm}{\hspace*{\fill}$\triangle$\vspace{1ex}}

\newcommand{\Z}{\mathbb{Z}}

\newcommand{\F}{\mathbb{F}}
\newcommand{\Fq}{\F_q}

\newcommand{\B}{\mathcal{B}}

\newcommand{\FF}{\mathbb{F}}

\newcommand{\RR}{\mathbb{R}}

\newcommand{\Zn}{\Z/ n\Z}

\newcommand{\fF}{\mathfrak{F}}

\newcommand{\bb}{{\bf b}}
\newcommand{\bc}{{\bf c}}
\newcommand{\bd}{{\bf d}}

\newcommand{\bv}{{\bf v}}
\newcommand{\bw}{{\bf w}}

\newcommand{\Tr}{\operatorname{Tr}}

\newcommand{\wq}{\ensuremath{w_q}}

\newcommand{\wqrel}{\ensuremath{w_q^{(s)}}}
\newcommand{\wrel}{\ensuremath{w_2^{(s)}}}

\newcommand{\amp}{\operatorname{amp}}

\newcommand{\recursion}[8]
{\def\arraystretch{1.8}
\begin{array}{ll}
#1 = & #2 \textrm{, and }\\
#3 = &\left\{\begin{array}{lc}
#4,&\textrm{if}\ #5\\
#6,&\textrm{if}\ #7 \\
\end{array}
\right. ,\\
& \textrm{ for } #8.
 \end{array}
}

\title{On Squares of Cyclic Codes}
\author{Ignacio Cascudo\thanks{Ignacio Cascudo is with the Department of Mathematical Sciences, Aalborg University, Denmark. Email:\texttt{ignacio@math.aau.dk}.
 The author was partially supported by the Danish Council for Independent Research, under grant no. DFF-4002-00367.
Accepted in Transactions on Information Theory, \url{https://ieeexplore.ieee.org/document/8451926/}. DOI 10.1109/TIT.2018.2867873. 
 Copyright~\copyright~2017 IEEE. Personal use of this material is permitted.  However, permission to use this material for any other purposes must be obtained from the IEEE by sending a request to pubs-permissions@ieee.org.
}}

\begin{document}
\maketitle

\begin{abstract}
The square $C^{*2}$ of a linear error correcting code $C$ is the linear code spanned by the component-wise products of every pair of (non-necessarily distinct) words in $C$. Squares of codes have gained attention for several applications mainly in the area of cryptography, and typically in those applications one is concerned about some of the parameters (dimension, minimum distance) of both $C^{*2}$ and $C$. In this paper, motivated mostly by the study of this problem in the case of linear codes defined over the binary field, squares of cyclic codes are considered. General results on the minimum distance of the squares of cyclic codes are obtained and constructions of cyclic codes $C$ with relatively large dimension of $C$ and minimum distance of the square $C^{*2}$ are discussed. In some cases, the constructions lead to codes $C$ such that both $C$ and $C^{*2}$ simultaneously have the largest possible minimum distances for their length and dimensions.

\end{abstract}

\section{Introduction}

The $m$-th Schur power $C^{*m}$ of a linear error correcting code $C$ is the linear code spanned by the component-wise products (or Schur products) of every tuple of $m$ (non-necessarily distinct) words in $C$. When $m=2$, we speak about the Schur square $C^{*2}$ of $C$. For the sake of conciseness we henceforth omit the name Schur and simply refer to squares, powers and products of codes. Powers and especially squares of codes play a relevant role in several recent results in cryptography and in particular in the area of secure multiparty computation (secure multiparty computation aims at solving the problem of how several mutually distrustful parties can jointly carry out computations involving private data known by some of them, without this private information being revealed to the other parties, see~\cite{CDN15} for more information about this area).
 In addition to this, the study of squares of codes is also useful for other applications such as the construction of bilinear multiplication algorithms in finite extensions of finite fields (through the notion of supercode introduced in \cite{STV92}) or the cryptanalysis of public key encryption schemes based on error correcting codes (see~\cite{COT17} and the references therein). Moreover, the notion of a square of a code is a special case of that of a component-wise product of two codes, which has been studied in connection to error-correcting pairs (also known as error locating pairs) for efficient error correction \cite{Koe92,Pel92}. As a consequence of this, properties of products and powers of codes have been analysed in recent years in works such as \cite{CCMZ15,Mir12,MZ15,Ran13AG,Ran13S,Ran15}. More information about general applications of squares of codes can be found in~\cite{Cas15, CCMZ15, CDN15,Ran15}.

In many of the applications above, we benefit from using codes $C$ such that, simultaneously, the minimum distance of $C^{*2}$ (denoted $d(C^{*2})$)  and the dimension of $C$, $\dim C$ are both large in relation to the length of $C$, and therefore the relationship between these parameters has been studied in recent works. However, most of the research so far has focused on the asymptotic setting, where the trade-offs between $d(C^{*2})$ and $\dim C$ are analyzed for a family of codes with lengths growing to infinity. In contrast there are not many results about how large these parameters can be when the length $n$ is fixed to be some particular value. The exception is the case where the $n$ is smaller than the size $q$ of the finite field over which $C$ is defined, since then it is known that Reed-Solomon codes give the best possible trade-offs (see below). However for some cryptographic applications, like the ones detailed below, using Reed-Solomon codes has the drawback that a large finite field needs to be used for accommodating a code of a given length. This increases both the computational and the communication complexity of the corresponding protocols in comparison to what one could achieve if long codes over smaller fields, especially the binary field, could be used instead.

Motivated by this problem, this paper considers cyclic codes which for the usual problem in coding theory where we want to optimize the trade-offs between $d(C)$ (instead of $d(C^{*2})$) and $\dim C$, give the best results for many values of the length $n$ (in the sense that for a given length and dimension, the minimum distance). The question we want to examine here is to what extent this is still true when we substitute $d(C)$ by $d(C^{*2})$.

The results in this paper strongly suggest this is the case. Indeed, constructions of cyclic codes where both $\dim C$ and $d(C^{*2})$ are relatively large with respect to $n$ are obtained and in particular, in the case of the binary finite field, the codes obtained by these constructions achieve the largest values for the minimum distance of their squares for several fixed values for their length and dimension.

\subsection{Related work}
A Singleton-like bound relating $\dim C$ and $d(C^{*2})$ was established in~\cite{Ran13S} and later the family of codes attaining this bound was characterized in~\cite{MZ15} (both works treat in fact the more general setting of products of codes). In particular, unless one of the two parameters ($\dim C$ or $d(C^{*2})$) is very restricted, Reed-Solomon codes are the only ones which can match this bound (see Section~\ref{sec:squares} for more information about these results).

However, as mentioned above, Reed-Solomon codes have the restriction that $n\leq q$. Therefore the asymptotic behaviour of families of squares of codes has been considered, where the finite field $\Fq$ is fixed and $n$ grows to infinity. The existence, over every finite field, of asymptotically good families\footnote{We say that a family of codes $\{C_i\}_{i\in \mathbb{N}}$ with lengths $n_i$ is asymptotically good if $n_i\rightarrow\infty$ when $i\rightarrow\infty$ and the limits $\lim_{i\rightarrow\infty}\dim C_i/n_i$ and  $\lim_{i\rightarrow\infty}d (C_i)/n_i$ exist and are strictly positive.} of codes whose squares also form an asymptotically good family was established in \cite{Ran13AG}. For small fields, this result requires a combination of an algebraic geometric construction over a sufficiently large (but constant) extension field and a special concatenation function to achieve a final construction over the small finite field. However, \cite{CCMZ15} showed that families of codes with such asymptotic properties are not very abundant, since choosing codes uniformly at random (among all codes of a prescribed dimension that grows linearly with the length) will, with high probability, not satisfy the desired properties.

Instead of considering the asymptotic setting, this paper focuses on specific values for the length of the code $C$; here the problem is that not many existing results that can be applied to, for example, the setting of linear binary codes with lengths, say, $n\leq 10000$. One option is to use Reed-Solomon codes over large enough extension fields paired with the concatenation technique in \cite{Ran13AG}. Reed-Muller codes are a family of binary codes for which it is relatively easy to determine the minimum distance of their squares.

Squares of cyclic codes have not been studied too much so far. In~\cite{Mir12} the square of a cyclic code, which is again cyclic, is described by relating its generator polynomial to that of the original code, and $\dim C^{*2}$ (and in some cases also $d(C^{*2})$) is computed for all cyclic codes of certain specific lengths and dimensions $\dim C$. Moreover, it is suggested that squares of cyclic codes have smaller dimensions than those of random codes. Other related results appeared in \cite{DK94}, who studied error-locating pairs for cyclic codes. While the application considered there is different, some of their intermediate results will be useful in the setting considered here too.

\subsection{Applications}
In order to justify the set of parameters this paper focuses on, some concrete applications are briefly mentioned now.
First, cyclic codes played a central role in a construction of a cryptographic tool known as additively homomorphic universally composable secure commitment schemes~\cite{CDD15}. This result requires binary codes $C$ with certain fixed $\dim C$ and $d(C)$ and, for those values, the shortest known codes are BCH codes, which are a family of cyclic codes. The concrete parameters that are considered in~\cite{CDD15}, when comparing the performance of their construction with previous alternatives, are $\dim C=256$ and $d(C)\geq 120$. However, the construction was further improved in~\cite{FJN16, CDD16} and it was shown that the same level of security can be achieved with a modified construction that only needs half the minimum distance. They consider the cases $\dim C=256$ and $d(C)\geq 40,60,80$ which achieve different levels of security. The complexity of the protocol depends on the length of the codes and it is advantageous for the construction that they are short. The constructions from~\cite{CDD15, FJN16, CDD16} attained several efficiency advantages with respect to prior work~\cite{DDGN14,GIKW14}, but lack one of the useful properties from~\cite{DDGN14} regarding verifiable commitment multiplication proofs. As suggested in \cite{Gia16}, one can recover this property by a small modification of the construction, but this requires replacing the requirement on $d(C)$ by the same one on $d(C^{*2})$. The question is then how much the length of the code (and consequently the complexity of the commitment protocol) needs to grow in order to accommodate this more stringent requirement.

Second, some of the currently best alternatives (in terms of communication complexity) for secure multiparty computation protocols for Boolean circuits were given in \cite{DZ13} (known as MiniMac) and its successor \cite{DNNR16} (which uses MiniMac as part of the construction). In MiniMac, a linear binary code $C$ is used, its role basically being to ensure that the parties behave honestly and do not change their private information in the middle of the computation. It is guaranteed that cheating players will be caught except with small probability. This probability depends on $d(C^{*2})$. On the other hand the dimension $\dim C$ (or more precisely the rate) will be related to the overhead in communication in the protocol.

Finally, linear codes with good squares can be used to construct strongly multiplicative secret sharing schemes, a notion introduced in \cite{CDM00}. More concretely, a linear code $C$ such that $d(C^{*2})\geq t+2$ and $d(C^{\bot})\geq t+2$, where $C^{\bot}$ denotes the dual code of $C$, gives raise to a $t$-strongly multiplicative secret sharing scheme, which in turn 
is enough to construct a multiparty computation protocol which is information-theoretically secure against $t$ corrupted players. Strongly multiplicative secret sharing schemes over small fields were studied in \cite{CC06, CCCX09, CCX14} in the asymptotic setting, but the constructions are algebraic geometric and present problems regarding the efficiency of computing the generator matrices for the codes. For relatively large values of $n$ one can be interested in using alternatives such as cyclic codes, for which a first step is understanding the behaviour of $d(C^{*2})$ relative to $\dim C$. However, it should be noted that in this paper we do not address the study of the dual of $C$, which is left for future work.

\subsection{Overview of the results}

The main goals of this paper are two: first, to give a description of the square $C^{*2}$ of a cyclic code $C$ that facilitates the task of finding tight lower bounds for $d(C^{*2})$; second, to exploit this description in order to find families of cyclic codes with simultaneously ``large'' (with respect to its length) values of $\dim C$ and $d(C^{*2})$, with special focus on binary codes and on the range of parameters which is interesting for the applications in~\cite{CDD15}.

The first goal is addressed with Theorem~\ref{thm:squarescyclic}, where it is shown that some observations from \cite{DK94} lead to a description of the generator polynomial that seems to present some advantages with respect to the one given in \cite{Mir12}. In particular it gives a direct description of the generating set of $C^{*2}$ in terms of the one for $C$, which allows for applying the BCH bound easily.

As for the second aim, several ways of choosing the generating sets of $C$ are suggested. The two first proposed constructions, described in Section~\ref{sec:preliminary}, only yield Reed-Solomon codes and punctured Reed-Muller codes.  A third approach is described in Section~\ref{sec:restricted}. It considers the case of codes of length $n=q^k-1$ and is based on the notion of restricted weights, which is introduced also in that section. Bounds for $d(C^{*2})$ are given and the dimension of $C$ is determined exactly by counting the number of walks of a given length in a certain graph. This construction is still parametrized by two integers, and in Section~\ref{sec:concrete} certain concrete values for these integers are fixed and explicit values for the lengths, dimensions and bounds for the minimum distances of binary codes and their squares are given.

It is seen that in some cases, the codes $C$ obtained satisfy the following two simultaneous features: $d(C)$ is the largest minimum distance possible for a code of length $n$ and dimension $\dim C$; and  $d(C^{*2})$ is the largest possible for a code of length $n$ and dimension $\dim C^{*2}$. In other cases, ``largest possible minimum distance" is replaced by ``largest minimum distance achieved by currently known codes (according to the code tables in \cite{Gra,Min})", since for those sets of parameters currently known lower and upper bounds for the minimum distance of codes do not coincide.

\section{Preliminaries}\label{sec:preliminaries}

Throughout this work, $q$ will be a power of a prime, and $\Fq$ will denote a finite field of $q$ elements. Let $n>0$ be a positive integer, and let $C$ be a linear code over $\Fq$ of length $n$, i.e., a $\Fq$-linear subspace of $\Fq^n$. Then $\dim C$ denotes the dimension of $C$ (its dimension as a vector space over $\Fq$); and the minimum distance of $C$, denoted as $d(C)$, is the smallest Hamming weight of a nonzero word in $C$.

Moreover given $\bv,\bw\in\Fq^n$, $\bv*\bw$ will denote their component-wise product as vectors in $\Fq^n$.

\subsection{Squares of codes}\label{sec:squares}

\begin{Definition}
Given two linear codes $C$ and $D$ over $\Fq$, their product $C*D$ is the linear code spanned over $\Fq$ by the set $\{\bc * \bd: \bc\in C, \bd \in D\}$.

The square of $C$ is the linear code $C^{*2}=C*C$, i.e., the linear code spanned over $\Fq$ by the set $\{\bc * \bc': \bc, \bc' \in C\}$.
\end{Definition}

Similarly one can recursively define the $m$-th power of $C$, for $m\geq 2$ as $C^{*m}=C^{*(m-1)}* C$. The primary focus of this paper are however the squares $C^{*2}$. Some relations between the dimensions and minimum distances of $C$ and $C^{*2}$ are given next. The proofs and generalizations of these results for higher powers and products of different codes can be found in~\cite{Ran15}.
\begin{Proposition}\label{prop:dimsquare}
The dimension of $C^{*2}$ satisfies $\dim C\leq \dim C^{*2}\leq \frac{(\dim C)\cdot(\dim C+1)}{2}.$ The minimum distance of $C^{*2}$ satisfies $d(C^{*2})\leq d(C)$.
\end{Proposition}

The two propositions above indicate that lower bounds for $\dim C$ and $d(C^{*2})$ will also be lower bounds for $\dim C^{*2}$ and $d(C)$ respectively, which is one of the reasons why we focus on the parameters $\dim C$ and $d(C^{*2})$. The following Singleton-like bound was shown in~\cite{Ran13S}. 
\begin{Proposition}[\cite{Ran13S}]\label{prop:singleton}
It holds that $$d(C^{*2})\leq \max\{1,n-2\dim C+2\}.$$
\end{Proposition} 

It was later shown in \cite{MZ15} that, unless either $\dim C$ or $d(C^{*2})$ is very small, the only codes that achieve the above bound are Reed-Solomon codes. More precisely,

\begin{Proposition}[\cite{MZ15}]\label{prop:MDS}
Suppose that $d(C^{*2})>1$. If $d(C^{*2})=n-2\dim C+2$, then $C$ is either a Reed-Solomon code or a direct
sum of self-dual codes, where self-duality is relative to a non-degenerate bilinear form which is not necessarily the standard inner product. Furthermore, if in addition $\dim C\geq 2$ and $d(C^{*2})\geq 3$, then $C$ is a Reed-Solomon code.
\end{Proposition} 

We consider now the squares of some known families of codes, starting by Reed-Solomon codes. 
Squares of Reed-Solomon codes are again Reed-Solomon codes. Given integers $0\leq m< n$, a finite field $\F$ of cardinality $|\F|\geq n$ and a vector $b=(b_1,b_2,\dots,b_n)\in\F^n$ of evaluation points under the condition that $b_i\neq b_j$ if $i\neq j$, the Reed-Solomon code $RS_{\F,b}(m,n)$ is defined as

$$RS_{\F,b}(m,n)=\{(f(b_1), f(b_2),\dots,f(b_n)): f\in\F[X], \deg f\leq m\}$$

and it is a code of dimension $m+1$ and minimum distance $n-m$. For any integer $u>0$ $$(RS_{\F,b}(m,n))^{*u}=RS_{\F,b}(um,n),$$ as long as $um<n$. Otherwise $$(RS_{\F,b}(m,n))^{*u}=RS_{\F,b}(n-1,n)=\F^n.$$

Similar arguments can be used for other families of evaluation codes: concretely, consider Reed-Muller codes, which consist of evaluations of multivariate polynomials.

\begin{Definition}
A binary Reed-Muller code of length $2^k$ and order $r$ (where $1\leq r\leq k$), for short a $RM(r,k)$ code, is a linear code of the form 

$$C=\{(f(\bb_1),\dots,f(\bb_{2^k}): f\in\FF_2[X_1,\dots,X_k], \deg f\leq r\}$$

where $\bb_1,\dots,\bb_{2^k}$ are all the distinct elements in $\FF_2^k$, in some order; here $\deg$ refers to the total degree of the $k$-variate polynomial $f$.
\end{Definition}

It is well known that the distance of a $RM(r,k)$ code is $2^{k-r}$ and its dimension is $\sum_{i=0}^r \binom{k}{i}$.

If $C$ is an $RM(r,k)$ code, then $C^{*2}$ is an $RM(2r,k)$ code (if $2r\leq k$; $C^{*2}=\F_2^{2^k}$ otherwise). Consequently:

\begin{Proposition}\label{prop:rm}
If $C$ is an $RM(r,k)$ code, then $\dim C=\sum_{i=0}^r \binom{k}{i}$, and if $2r\leq k$, then $d(C^{*2})=2^{k-2r}$.
\end{Proposition}

In spite of these observations, squaring is a quite ``destructive'' operation for most codes; indeed it was shown in \cite{CCMZ15} that for large enough $k$ and $n$, if a linear code $C$ is chosen uniformly at random among all codes of dimension $k$ and length $n$ then with high probability the dimension of $C^{*2}$ will be very close to the ``maximal possible dimension'' $\min \{n,k(k+1)/2\}$ (see \cite{CCMZ15} for the precise statements). This implies that, for  a random family of codes, with very high probability either the family itself or the family of their squares will be asymptotically bad.

On the other hand, a construction, over every finite field, of asymptotically good families of linear codes whose squares are also an asymptotically good family was shown in~\cite{Ran13AG}. In order to obtain this result for small finite fields,~\cite{Ran13AG} needs to use families of algebraic geometric codes over a fixed extension field together with a specially crafted map that is used for concatenation, so that the resulting construction is over the desired small field. Since this concatenation is also relevant for comparison in the non-asymptotic setting, the concrete result is stated next.
\begin{Proposition}\cite[Corollary 12]{Ran13AG}\label{prop:concatenate}
Let $C$ be a linear code of length $n$ over the finite field $\FF_{q^{2s+1}}$. Then there exists a linear code $\phi(C)$ of length $n(s+1)(2s+1)$ over $\FF_q$ such that $\dim(\phi(C))=(2s+1)\dim C$ and $d(\phi(C)^{*2})\geq d(C^{*(1+q^s)})$.
\end{Proposition} 

In particular, by setting $C$ to be a Reed-Solomon code of length $q^{2s+1}$ and dimension $m+1$, we have 
\begin{Corollary}\label{cor:RSconcatenate}
For any integers $m,s>0$ such that $m<q^{(2s+1)}/(q^s+1)$, there exists a linear code $D$ over $\Fq$ of length $(s+1)(2s+1)q^{2s+1}$ with  $\dim D=(2s+1)(m+1)$ and $d(D^{*2})\geq q^{2s+1}-m(q^s+1)$.
\end{Corollary}

We point out now some transformations that allow to  obtain new codes from given ones and how they affect the square operation and their parameters. First given a linear code $C$, one can consider the code $D$ with codewords of the form $(\bc,\bc,\dots,\bc)$, $\bc\in C$.

\begin{Proposition}\label{prop:initial}
Given a linear code $C$ of length $n$ over $\Fq$, and an integer $m>0$ there exists another code $D$ over $\Fq$ with length $mn$, $\dim D=\dim C$, and $d(D^{*2})=m d(C^{*2})$. 

In particular, for every finite field $\Fq$ and any integers $n,m$ there exists a linear code $D$ over $\Fq$ of length $nm$ dimension $\dim D=n$ and minimum distance $d(D)=m$.  
\end{Proposition}
The last statement is obtained by setting $C=\Fq^n$.
Next, the well-known puncturing and shortening operations yield the following result.

\begin{Proposition}\label{prop:puncshor}
Let $C$ be a linear code of length $n$. For any $a,b$ non-negative integers with $a+b< n$, and $b<d(C)$, there exists a linear code $D$ of length $n-a-b$ and such that $\dim D\geq \dim C-a$ and $d(D^{*2})\geq d(C^{*2})-b$.
\end{Proposition}

\subsection{Cyclic codes}
From now on it will always be assumed that $n$ is coprime with $q$. There are a few different ways of defining a cyclic code, and it will be useful to consider two of them. The most common one is as follows. Consider the $\Fq$-vector space $R=\Fq[X]/(X^n-1)$. Since $R$ has dimension $n$ it is isomorphic as a $\Fq$-vector space to $\Fq^n$ and an isomorphism $\iota:\Fq^n\rightarrow R$ is given by $$(c_0,c_1,\dots,c_{n-1})\mapsto c(X)+\langle X^n-1 \rangle$$ where $c(X):=\sum_{i=0}^{n-1} c_i X^i$.

$R$ is a ring with the product operation induced by the usual product of polynomials in $\Fq[X]$.  From now on, the elements in $R$ are identified with polynomials in $\Fq[X]$ of degree at most $n-1$, since every class in $R$ has exactly one representative of that form. 

\begin{Definition}
Let $g\in\Fq[X]$ be a polynomial dividing $X^n-1$. The cyclic code generated by $g$ is the ideal generated by $g$ in $R$.
\end{Definition}

\begin{Lemma}\label{lem:dimensioncyclic}
The dimension of the cyclic code $C$ generated by $g$ is $n-\deg g$, since
$$C=\{g\cdot h\ |\ h\in\Fq[X],\ \deg h\leq n-\deg g-1\}.$$
\end{Lemma}
Let $\beta$ be a primitive $n$-root of unity in an algebraic closure of $\Fq$, i.e., $\beta^n=1$ but $\beta^k\neq 1$ for $1\leq k\leq n$. Let $\fF=\Fq(\beta)$ be the smallest field containing $\beta$ and $\Fq$. $\fF$ is in fact a finite field $\FF_{q^r}$ of $q^r$ elements, where $n$ divides $q^r-1$.

Since $g$ divides $X^n-1$, all roots of $g$ are of the form $\beta^j$, for some $j\in\{0,\dots,n-1\}$. As a matter of fact, since $\beta$ is a $n$-root of unity, we can also define the notation $\beta^j$ for $j\in\Z/n\Z$.

\begin{Definition}
We call $J:=\{j\in\Zn: g(\beta^j)=0\}$ and $I:=\{j\in \Zn: g(\beta^j)\neq 0\}$ respectively the defining and generating sets of the cyclic code $C$ generated by $g$.
\end{Definition}

Note that $g=\prod_{j\in J} (X-\beta^j)= (X^n-1)/\prod_{i\in I} (X-\beta^i)$ and hence $\dim C=|I|$, where $|I|$ denotes the cardinality of $I$. Since $g$ is in $\Fq[X]$, whenever $\gamma$ is a root of $g$, $\gamma^q$ is a root too, and hence there are some restrictions to $J$ and $I$:

\begin{Definition}
Let $u\in \Z/n\Z$. The $q$-cyclotomic coset of $u$ is the set $[u]:=\{uq^j: j\geq 0\}\subseteq \Zn$ (where the products are understood to be in $\Zn$). 
\end{Definition}
\begin{Lemma}
Both $I$ and $J$ are unions of $q$-cyclotomic cosets.
\end{Lemma}

A key result in the theory of cyclic codes is the following

\begin{Proposition}[BCH bound]\label{prop:bch}
Suppose that $c,d\in\Zn$ are such that $\{c,c+1,\dots,c+d-2\}\subseteq J$. Then the minimum distance of $C$ is at least $d$. 
\end{Proposition}

This motivates the definition of BCH code.
\begin{Definition}
A BCH code of designed distance $d$ is a cyclic code with generator polynomial $g=lcm\{m_j: j\in\{c,c+1,\dots,c+d-2\}\}$, where $m_j$ is the minimal polynomial of $\beta^j$. That is, the defining set $J$ is the union of the cyclotomic cosets containing the elements $c,c+1,\dots,c+d-2$.
\end{Definition}

\begin{Lemma}\label{lem:bch}
The minimum distance of a BCH code of designed distance $d$ is at least $d$. Its dimension is at least $n-m(d-1)$, where $m$ is the smallest integer such that $n|(q^m-1)$. If $q=2$ and $c=1$, then its dimension is at least $n-md/2$.
\end{Lemma}

The dual of a cyclic code is another cyclic code. In fact, the following holds.
\begin{Definition}
For $I\subseteq \Zn$, $-I$ denotes the set $$-I:=\{-i: i\in I\}\subseteq \Zn.$$ 
\end{Definition}
\begin{Lemma}
Let $C$ be the cyclic code generated by $g$ and let $h=\frac{X^n-1}{g(X)}:=\sum_{i=0}^{|I|} h_iX^i$. Then

\begin{itemize}
\item The dual $C^{\bot}$ of $C$ is the cyclic code generated by the polynomial $$h^{[-1]}:=\sum_{i=0}^{|I|} h_{|I|-i}X^i.$$

\item Let $J$ and $I$ be the defining and generating sets of $C$ and let $J^*$ and $I^*$ the defining and generating sets of $C^{\bot}$. Then $J^*=-I$ and $I^*=-J$.
\end{itemize}
\end{Lemma}

It is more useful for the problem in hand to consider the following alternative description of cyclic codes as a subfield subcode of an evaluation code over the field $\fF=\FF_{q^r}$. We now follow the notation from~\cite{Bie02}.

\begin{Definition}\label{def:bier}
For a set $M\subseteq\{1,\dots,n-1\}$, we denote by ${\cal P}(M)$ the $\fF$-span of the monomials $X^i$, $i\in M$, i.e.,
$${\cal P}(M):=\{\sum_{i\in M}f_i X^i: f_i\in \fF\}.$$

In addition, let ${\cal B}(M)$ denote the $\fF$-vector space $${\cal B}(M):= \{(f(1), f(\beta),\dots,f(\beta^{n-1})): f\in {\cal P}(M) \}\subseteq\fF^n.$$
Finally as it is usual, for a set $V\subseteq\fF^n$, denote $V|_{\Fq}=V\cap\Fq^n$.
 
\end{Definition}

\begin{Lemma}\label{lem:eval}
Let $C$ be the cyclic code generated by $$g=\frac{X^n-1}{\Pi_{i\in I} (X-\beta^i)}\in\Fq[X].$$ Then $$C={\cal B}(-I)|_{\Fq}.$$
\end{Lemma}
\begin{Proof}
In \cite[Section 3]{Bie02}, it is established that $C=\Tr({\cal B}(J))^{\bot}$, where

$$\Tr(V):=\{(\Tr(v_1),\Tr(v_2),\dots,\Tr(v_n)): (v_1,v_2,\dots,v_n)\in V\}.$$

 and $\Tr$ denotes the trace from $\fF$ to $\Fq$. Given that $I$ is the complement of $J$ in $\{1,\dots,n\}$ and that $I$ is a union of cyclotomic sets, \cite[Theorem 6]{Bie02} states that $\Tr({\cal B}(J))^{\bot}={\cal B}(-I)|_{\Fq}$.
\end{Proof}

\section{Squares of cyclic codes}\label{sec:squarescyclic}

Consider the description of a cyclic code as an ideal in $R$. Given the identification between $R$ and $\Fq^n$, we can talk about the coordinatewise product of elements in $R$; more precisely, let $$h*h':=\sum_{i=0}^{n-1} h_ih'_iX^i.$$

Note that given a cyclic code $C$ with generator polynomial $g$, $C^{*2}$ consists of all elements of the form $$\sum_i \lambda_i (a_i\cdot g)*(b_i\cdot g) $$ where $\lambda_i\in\Fq$ and $a_i,b_i\in R$ are such that $\deg a_i, \deg b_i\leq n-\deg g-1$. Furthermore it was observed in \cite{Mir12} that
\begin{Lemma}[\cite{Mir12}]
$C^{*2}$ is a cyclic code with generator polynomial$$ g'=\gcd\{g*g,g*(g\cdot X),g*(g\cdot X^2),\dots, g*(g\cdot X^{n-\deg g-1})\}$$
\end{Lemma}

However, this description is not too easy to work with. Instead, it seems much more useful to 
use the interpretation of a cyclic code as an evaluation code, given by Lemma~\ref{lem:eval}, and then argue about the squares similarly to how it is done in the case of Reed Solomon codes. Consider the following definition.
\begin{Definition}
For subsets $A,B\subseteq\Zn$, define $A+B:=\{i+j: i\in A, j\in B\}\subseteq\Zn$.
\end{Definition}

Now note that if $f$ and $f'\in {\cal P}{(-I)}$ (notation as in Definition~\ref{def:bier}) then $f\cdot f'(\mod X^n-1)$ is in ${\cal P}(-(I+I))$. Hence ${\cal B}(-I)^{*2}= {\cal B}(-(I+I))$.

Let $C={\cal B}(-I)|_{\Fq}$, which by Lemma~\ref{lem:eval} is a cyclic code. It will be shown now that $C^{*2}={\cal B}(-I)^{*2}|_{\Fq}= {\cal B}(-(I+I))|_{\Fq}$, which is a special case of one of the observations in \cite{DK94} (stated there for a more general result for the product of two, non-necessarily equal, cyclic codes). Note that this observation is not immediate, as the operations of squaring and taking subfield subcodes do not commute in general.

\begin{Theorem}\label{thm:squarescyclic}
If $C={\cal B}(-I)|_{\Fq}$, then $C^{*2}={\cal B}(-(I+I))|_{\Fq}$.

In other words, if C is a cyclic code  generated by the polynomial $$g=\frac{X^n-1}{\prod_{i\in I}(X-\beta^i)},$$ then $C^{*2}$ is a cyclic code with generator polynomial $$g'=\frac{X^n-1}{\prod_{\ell\in I+I}(X-\beta^\ell)}.$$
\end{Theorem}

\begin{Proof}
We first note that extension by scalars of the extension field $\fF$ gives back ${\cal B}(-I)$, i.e.  ${\cal B}(-I)=\fF\otimes C$. Indeed $\fF\otimes C\subseteq {\cal B}(-I)$ is obvious (since ${\cal B}(-I)\subseteq \fF^n$ is a $\fF$-vector space containing $C$). Moreover $C$ is a vector space over $\Fq$ of dimension $|I|$ by Lemma~\ref{lem:eval}. Hence $\fF\otimes C$ is a $\fF$-vector space of dimension $|I|$. Moreover, it is immediate that ${\cal B}(-I)$ also has dimension $|I|$ as a $\fF$-vector space, so both spaces must be equal.
Now we can apply \cite[Lemma 2.23(iii)]{Ran15} which shows that extension by scalars always commutes with squaring. That is, $(\fF\otimes C)^{*2}=\fF\otimes C^{*2}$. Therefore $C^{*2}=(\fF\otimes C^{*2})|_{\Fq}=(\fF\otimes C)^{*2}|_{\Fq}={\cal B}(-I)^{*2}|_{\Fq}={\cal B}(-(I+I))|_{\Fq}$.

\end{Proof}

\begin{Remark}
As an aside note that, when $D$ is a linear code over a finite extension $\fF$ of $\Fq$, and $C=D|_{\Fq}$, we can always state that $C^{*2}\subseteq D^{*2}|_{\Fq}$ (even though equality does not hold in general, as mentioned above) and hence $d(C^{*2})\geq d(D^{*2}|_{\Fq})$. 
\end{Remark}

The discussion above implies that the BCH bound (Proposition~\ref{prop:bch}) can be used on the square of the cyclic code. To simplify the exposition, the following notation is introduced.

\begin{Definition}
Let $A\subseteq \Zn$ be a nonempty set. Its amplitude $\amp A$ is

$$\amp A:= \min \{i\in\{1,\dots,n\}: \exists c\in\Zn \textrm{ such that } A\subseteq \{c,c+1,\dots,c+i-1\} \}$$

(where sums are understood to be in $\Zn$). That is, $\amp A$ is the size of the smallest set of consecutive elements in $\Zn$ that contains $A$.
\end{Definition}

\begin{Remark}
Note that 
\begin{itemize}
\item $\amp A\leq 1+\max A$ where $\max A$ denote the largest element of $A$ when $\Zn$ is identified with the set of integers $\{0,\dots,n-1\}$. This is because $A\subseteq \{0,\dots,\max A\}$. 

\item $n-\amp A$ is the size of the largest set of consecutive elements that do not belong to $A$, i.e., the largest set of consecutive elements contained in $A^c$. 

\item It is then a direct consequence of Proposition~\ref{prop:bch} that the minimum distance of a cyclic code $C$ satisfies $d(C)\geq n-\amp I+1$ (remember $I^c=J$). 
\end{itemize}
\end{Remark}

\begin{Theorem}\label{thm:bounds} Let $C$ be a cyclic code of length $n$ with generator polynomial $g=(X^n-1)/f(X)$ where $f=\prod_{i\in I}(X-\beta^i)$.

 Then 
 \begin{itemize}
 \item $\dim C=|I|$ and $\dim C^{*2}=|I+I|$.
 \item $d(C)\geq n-\amp I+1$ and $d (C^{*2})\geq n- \amp (I+I)+1$.
\end{itemize}
\end{Theorem}

Thus, finding $I\subseteq \{0,\dots,n-1\}$ such that $I$ is a union of cyclotomic sets, and $|I|$ is large but $\amp(I+I)$ is relatively small will yield codes $C$ such that $\dim C$ and $d(C^{*2})$ are simultaneously large.

\section{Some preliminary constructions}\label{sec:preliminary}
In this section some natural approaches towards constructing the index sets $I$ are analysed. However, the two approaches in this section will lead respectively to Reed-Solomon and generalized Reed-Muller codes, whose squares are well understood, as  discussed in Section~\ref{sec:squares}.
Nevertheless, they will also provide useful intuitions for the more involved techniques presented in Section~\ref{sec:restricted}, so it is still interesting to elaborate on them here.

The first approach consists in taking the generator set $I$ to be the union of all cyclotomic sets that are entirely contained in $\{0,\dots,t\}$ for some integer $t<n/2$. The idea is that $I+I$ is then contained in $\{0,\dots,2t\}$ and therefore its amplitude is (at most) $2t+1$, which gives a lower bound $d (C^{*2})\geq n- 2t$.
Note that the complement of $I$, the defining set $J$, is the smallest union of cyclotomic sets containing $\{t+1,\dots,n-1\}$. Hence the generator polynomial is $g:=\textrm{mcm}(m_{t+1},m_{t+2},\dots,m_{n-1})$ where $m_i$ is the minimal polynomial in $\Fq[X]$ of $\beta^i$. In other words, the code $C$ is a BCH code of designed distance $n-t$.

This immediately suggests the following consequence:

\begin{Theorem}\label{thm:first}
Let $t,n$ be positive integers and let $k$ be the smallest integer with $n| (q^k-1)$.
There exists a $\Fq$-linear code $C$, of length $n$ such that
\begin{itemize}
\item $\dim C\geq \max\{1, n-(n-t-1)k\}$,
\item $d(C)\geq n-t$ and
\item $d(C^{*2})\geq n-2t$.
\end{itemize}
If in addition $q=2$, then $\dim C\geq \max\{1, n-\frac{(n-t)k}{2}\}.$
\end{Theorem}

\begin{Proof}
Take as generator of $C$ the polynomial $g:=\textrm{mcm}(m_{t+1},m_{t+2},\dots,m_{n-1})$ where $m_i$ is the minimal polynomial in $\Fq[X]$ of $\beta^i$.
The statements about the distance of $C$ and $C^{*2}$ follow from Main Theorem \ref{thm:bounds} and the fact that the amplitudes of $I$ and $I+I$ are at most $t+1$ and $2t+1$ respectively. On the other hand, the estimates about the dimension are as in Lemma~\ref{lem:bch}.

\end{Proof}

Unfortunately, the result above cannot be used to ensure that $\dim C>1$ and $d(C^{*2})>1$ simultaneously, unless in the case where $k=1$. However, in that case $C$ is just a Reed Solomon code over $\Fq$.

A different idea will be considered next. Given that the set $I$ generating the code needs to be a union of cyclotomic cosets, we can think of associating to each integer a quantity which is invariant within a cyclotomic coset and at the same time can be ``controlled'' to a certain extent when two integers are summed.

We will from now on consider the case $n=q^k-1$. Then we can use as invariant the $q$-ary weight, defined next.
 
\begin{Definition}
The $q$-ary representation of an element $t\in\Zn$ is the unique vector $$(t_{k-1},t_{k-2},\dots,t_0)_q\in\{0,\dots,q-1\}^k$$ such that $t=\sum_{i=0}^{k-1} t_iq^i$. The $q$-ary weight of $t$ is defined as $\wq(t)=\sum_{i=0}^{k-1} t_i$. 
\end{Definition}

\begin{Lemma}\label{lem:weights}
Let $n=q^k-1$. Let $a,b\in\Zn$. Then:
\begin{itemize}
\item $\wq(q^j a)=\wq(a)$ for any $j\geq 0$, i.e., all elements in the same $q$-cyclotomic coset have the same $q$-ary weight.
\item $\wq(a+b)\leq \wq(a)+\wq(b).$
\end{itemize}

\end{Lemma}

\begin{Proof}
The first part of the lemma comes from the fact that multiplying by $q$ simply induces a cyclic shift on the $q$-ary representation of an element of $\Zn$, because $n$ is of the form $q^k-1$. For the second part, let $a=\sum_{i=0}^{k-1} a_iq^i, b=\sum_{i=0}^{k-1} b_iq^i$. If $0\leq a_i+b_i\leq q-1$ for all $i$ (i.e., if there are no carries in the sum), then $\wq(a+b)=\wq(a)+\wq(b)$. Otherwise, whenever there is a carry, the weight will decrease by $q-1$.

\end{Proof}

The first part of the lemma implies that we can talk about the $q$-ary weight of a $q$-cyclotomic set (which is the $q$-ary weight of any of its elements). The second part leads to the following:

\begin{Proposition}\label{prop:weightdoubles}
If $I$ is the union of all cyclotomic sets whose $q$-ary weights are at most\footnote{There is nothing special about choosing a bound of the form $(q-1)h$. It just leads to a simpler expression for the square distance, but the result can be generalized to other values which are not divisible by $q-1$.} $(q-1)h$, for some integer $h\geq 1$, then $I+I$ is a union of cyclotomic sets of weight at most $2(q-1)h$, and moreover $\amp(I+I)\leq 1+q^k-q^{k-2h}$.
\end{Proposition}

\begin{Proof}
The first part of the proposition follows directly from Lemma~\ref{lem:weights}. For the second part, note $q^k-q^{k-2h}=(q-1)\sum_{\ell=1}^{2h} q^{k-\ell}$  has $q$-ary representation $(\overbrace{q-1,q-1,\dots,q-1}^{2h \textrm{ times}},0,0,\dots,0)_q$.
Hence it is obviously the largest integer in $\{0,\dots,n-1\}$ of weight at most $2(q-1)h$. On the other hand, this integer can be written as the sum of two integers from $I$, namely $$(q-1)\sum_{\ell=1}^{2h} q^{k-\ell}= (q-1)\sum_{\ell=1}^{h} q^{k-\ell}+(q-1)\sum_{\ell=h+1}^{2h} q^{k-\ell},$$ so it is indeed in $I+I$. We have shown $\max (I+I)=q^k-q^{k-2h}$, and hence $I+I\subseteq \{0,\dots,q^k-q^{k-2h}\}$ so the amplitude of $I+I$ is at most $1+q^k-q^{k-2h}$. 
\end{Proof}

\begin{Corollary}\label{cor:weightconstruction}
Let $n=q^k-1$ and let $C$ be the cyclic code generated by the polynomial $g=(X^n-1)/f(X)$, where $f=\prod_{i\in I} (X-\beta^i)$ and $I=\{i:\wq(i)\leq (q-1)h\}$.
We have $d(C^{*2})\geq q^{k-2h}-1$. 

\end{Corollary}

\begin{Proof}
It follows from Main Theorem~\ref{thm:bounds} and Proposition~\ref{prop:weightdoubles}. 

\end{Proof}

Nevertheless, these cyclic codes are in fact generalized Reed-Muller codes punctured in one position, as it is shown next.

\begin{Remark}\label{rem:inv}
By setting $\alpha=\beta^{n-1}$ (which is again a primitive $n$-th root of unity), and noticing that $\wq(n-i)=(q-1)k-\wq(i)$, it is easy to see that $g=\prod_{j\in J'} (X-\alpha^j)$ and $1\leq \wq(j)\leq (q-1)k-(q-1)h$ for all $j\in J'$. 
\end{Remark}

\begin{Proposition}\label{prop:grm}
In the conditions of Corollary~\ref{cor:weightconstruction}, $C$ is equivalent to a generalized Reed-Muller code of length $q^k$ and order $(q-1)h$ punctured in one position.
\end{Proposition}

\begin{Proof}
By Remark~\ref{rem:inv}, the polynomial $g$ can be written as $g=\prod_{1\leq \wq(j)\leq (q-1)k-(q-1)h}(X-\alpha^j).$
Now let $C'$ be the cyclic code generated by the polynomial $$g'=(X-1)g= \prod_{0\leq \wq(j)\leq (q-1)k-(q-1)h}(X-\alpha^j)$$ and let $D$ be the code of length $q^k$ spanned by the vectors $\{(c',0): c'\in C'\}\cup\{(1,\dots,1)\}\subseteq \FF_q^{n+1}$. It is known~\cite{Lint82} that $D$ is equivalent to the Reed Muller code of length $q^k$ and order $(q-1)h$. Puncturing this code in the last position we obtain the code spanned by $C'\cup\{(1,\dots,1)\}\subseteq \FF_q^n$. It is easy to see that this code is $C$, since $\dim C=\dim C'+1$, $C$ contains $C'$, and $C$ contains the vector $(1,\dots,1)\in\FF_q^n$ (because $1+X+\dots+X^{n-1}=(X^{n}-1)/(X-1)$ is clearly a multiple of $g$).

\end{Proof}

\section{Construction of codes based on restricted weights}\label{sec:restricted}
In this section, a modification of the second approach from the previous section is suggested. As in the last part of the previous section, the length of the codes will be $n=q^k-1$ for some $k$.
The modification consists on replacing the notion of $q$-ary weight by a 
 notion of weight which is defined next. 

\begin{Definition}
 Let $t\in\{0,\dots,n-1\}$ with $q$-ary representation $(t_{k-1},t_{k-2},\dots,t_0)_q$, and let $1\leq s\leq k$. 
 
 The $s$-restricted binary weight of $t$ is defined as $\wqrel(t)=\max_{i\in\{0,\dots,k-1\}} \sum_{j=0}^{s-1} t_{i+j}$, where the sums $i+j$ are considered modulo $k$.  
\end{Definition}
Therefore, the $s$-restricted weight of $t$ is the maximum weight of a substring of $s$ consecutive digits in the q-ary representation of $t$. Here ``consecutive'' is also meant cyclically, and hence it is clear that this notion is an invariant of a cyclotomic coset, i.e., $\wqrel(q^it\mod n)=\wqrel(t)$ for any $i\geq 0$. Thus, we can speak of the $s$-restricted binary weight of a cyclotomic coset.

\begin{Remark}
If $s=k$, then $\wqrel(t)=\wq(t)$.
\end{Remark}

Moreover the notion of restricted weight also satisfies the subadditivity property.

\begin{Proposition}\label{prop:srw}
Let $t, u\in\{0,\dots,n-1\}$. Let $v:=t+u \mod n$. Then $\wqrel(v)\leq \wqrel(t)+\wqrel(u)$.
\end{Proposition}

The proof of this result is somewhat tedious and it is therefore deferred to the appendix.

In view of the proposition above, it is clear that if we take $I$ to contain only elements of $s$-restricted weight at most $m$, then all elements in $I+I$ will have $s$-restricted weight at most $2m$. This motivates the following definitions and results.

\begin{Definition}
Recall that $n=q^k-1$. We denote:
$$W_{k,s,m}:= \{j\in \{0,\dots,n-1\}: \wqrel(j)\leq m\},$$
$$N_{k,s,m}:=|W_{k,s,m}| \textrm{  and  }$$
$$B_{k,s,m}:= \max\ W_{k,s,m}.$$

\end{Definition}

Here, the finite field size $q$ has been omitted from the definitions for simplicity of notation, but note that these numbers do depend on which $q$ we are considering

\begin{Proposition}\label{prop:bsm}
We have $\amp(W_{k,s,m})\leq 1+B_{k,s,m}$.
Furthermore if $m\leq \frac{s-1}{2}$, then $W_{k,s,m}+W_{k,s,m}\subseteq W_{k,s,2m}$ and consequently $\amp(W_{k,s,m}+W_{k,s,m})\leq 1+B_{k,s,2m}$.

\end{Proposition}

\begin{Proof}
This is straightforward from the definitions above and Proposition~\ref{prop:srw}.
\end{Proof}

The equality $W_{k,s,m}+W_{k,s,m}=W_{k,s,2m}$ does not necessarily hold; for example, for $q=2$, it holds on the one hand that $W_{5,3,1}=\{0,1,2,4,8,16\}$, the set of all binary strings of length $5$ and weights $0$ and $1$ (indeed, given any string of weight at least $2$, one can find $3$ cyclically consecutive positions containing two $1$'s, so it cannot belong to $W_{5,3,1}$). By Lemma~\ref{lem:weights}, every element in $W_{5,3,1}+W_{5,3,1}$ has weight at most $2$. On the other hand, since the binary representation of $26$ is $11010$, then $w_2^{(3)}(26)=2$. Therefore we have $26\in W_{5,3,2}\setminus(W_{5,3,1}+W_{5,3,1})$.

This observation in fact provides a tighter bound for $\amp(W_{k,s,m}+W_{k,s,m})$, as follows.

\begin{Proposition}\label{prop:wksm}
Let $t\in W_{k,s,m}$. Then $w_q(t)\leq \left\lfloor\frac{mk}{s}\right\rfloor $.
\end{Proposition}

\begin{Proof}
The $q$-ary representation of $t$ contains $k$ different substrings of $s$ cyclically consecutive positions, and each position belongs to $s$ of these strings. Hence the sum $S$ of the weights of these strings is exactly $S=s w_q(t)$. On the other hand, each of these strings has weight at most $m$, and hence $S\leq km$. Hence $w_q(t)\leq  \frac{mk}{s}$ and the result follows from the fact that $w_q(t)$ is an integer.
\end{Proof}

\begin{Corollary}\label{cor:w2bound}
Let $t\in W_{k,s,m}+W_{k,s,m}$. Then $w_q(t)\leq 2\left\lfloor\frac{mk}{s}\right\rfloor $.
\end{Corollary}
Note that solely from the fact that $t\in W_{k,s,2m}$, one can only guarantee that $w_q(t)\leq \left\lfloor\frac{2mk}{s}\right\rfloor$. This may be larger than $2\left\lfloor\frac{mk}{s}\right\rfloor$, as it happens in the example above.

\begin{Definition}
Let $$\widehat{B}_{k,s,2m}:= \max\left\{t\in W_{k,s,2m}: w_q(t)\leq 2\left\lfloor\frac{mk}{s}\right\rfloor\right\}.$$
\end{Definition}

\begin{Theorem}\label{thm:codesm}
let $C$ be the cyclic code generated by the polynomial $g=(X^n-1)/f(X)$, where $f=\prod_{i\in W_{k,s,m}} (X-\beta^i)$.
Then
\begin{itemize}
\item $\dim C=N_{k,s,m}.$
\item $d(C)\geq n-B_{k,s,m}.$
\item $d(C^{*2})\geq n-\widehat{B}_{k,s,2m}.$
\end{itemize}
\end{Theorem}

A slight variation of this result can be obtained if the index $0$ is removed from $W_{k,s,m}$. Let $I:=W_{k,s,m}\setminus\{0\}$. Then obviously $|I|=N_{k,s,m}-1$. On the other hand, if $2\left\lfloor\frac{mk}{s}\right\rfloor<k$, then $0\notin I+I$. In these conditions, $I+I\subseteq\{1,\dots,\widehat{B}_{k,s,2m}\}$. Hence

\begin{Theorem}\label{thm:codesmz}
let $C$ be the cyclic code generated by the polynomial $g=(X^n-1)/f(X)$, where $f=\prod_{i\in W_{k,s,m}\setminus\{0\}} (X-\beta^i)$. In addition, assume $2\left\lfloor\frac{mk}{s}\right\rfloor<k$.
Then
\begin{itemize}
\item $\dim C=N_{k,s,m}-1.$
\item $d(C)\geq n-B_{k,s,m}+1.$
\item $d(C^{*2})\geq n-\widehat{B}_{k,s,2m}+1.$
\end{itemize}
\end{Theorem}

The rest of this section is devoted to analyse the numbers $B_{k,s,m}$, $\widehat{B}_{k,s,2m}$, and $N_{k,s,m}$.

\subsection{Bounds for the distance of the codes and their squares.}

In order to calculate $B_{k,s,m}$ and $\widehat{B}_{k,s,2m}$ one can simply consider their $q$-ary representations and determine their digits one by one, going from the highest order digit to the lowest one and assigning, at each step, the largest value in $\{0,\dots,q-1\}$ that is consistent with the conditions on the weights.

As an example consider the case $q=2$. Assuming $n\geq s$, the binary representation of $B_{k,s,m}$ will begin with $\lfloor k/s \rfloor$ blocks of the form $11...100...0$ ($m$ ones and $s-m$ zeros). The remaining $k-s\lfloor k/s \rfloor<s$ positions should contain 1's until the last $s-m$ positions are reached: these must in any case be all zero, because of the fact that the first $m$ positions are one, and that the definition of restricted weight considers any set of $s$ consecutive positions cyclically.
The binary representation of $\widehat{B}_{k,s,2m}$ is obtained from the representation of ${B}_{k,s,2m}$ by swapping the 1's in the lowest order coordinates to $0$'s until the weight is at most $2\left\lfloor\frac{mk}{s}\right\rfloor $. By proceeding in this manner one obtains the following formulas.

\begin{Lemma}\label{lem: bsm}
Let $q=2$ and $1\leq m\leq s-1$. Then $$B_{k,s,m}=\sum_{i=0}^{\lfloor \frac{k}{s}\rfloor -1}\sum_{j=1}^{m} 2^{k-is-j}+\sum_{i=s-m}^{k-s\lfloor \frac{k}{s}\rfloor-1} 2^i $$ 

Let $1\leq m\leq \frac{s-1}{2}$. Then $$\widehat{B}_{k,s,2m}=\sum_{i=0}^{\lfloor \frac{k}{s}\rfloor -1}\sum_{j=1}^{2m} 2^{k-is-j}+\sum_{i=u}^{k-s\lfloor \frac{k}{s}\rfloor-1} 2^i $$

where 
$u=\max\{s-2m,(k-s\lfloor \frac{k}{s}\rfloor)-(2\left\lfloor\frac{mk}{s}\right\rfloor-2m\left\lfloor\frac{k}{s}\right\rfloor)\}$.
\end{Lemma}

\begin{Remark}
In particular, for $q=2$ we have $\widehat{B}_{k,s,2m}< 2^k-2^{k-2m-1}$ and therefore, if $C$ is defined as in Theorem~\ref{thm:codesm}, then $d(C^{*2})\geq 2^{k-2m-1}$.
\end{Remark}

\subsection{Determining the dimension of the codes}\label{sec:dimension}

In this section, a recurrence formula for the numbers $N_{k,s,m}$ with respect to $k$ will be found.
Remember $N_{k,s,m}$ equals the number of strings in $\{0,\dots,q-1\}^k$ such that every sequence of $s$ consecutive positions of the string and of its cyclic shifts contains at most $m$ ones.

In the case $q=2$ and if we remove the cyclic condition (meaning for example that in the case $k=4$, $s=3$, $m=2$, the string 1011 would be included in the counting while its cyclic shift 1110 would not) a solution for certain parameters of $s, m$ can be found in the online encyclopedia of integer sequences~\cite{OEIS}. More concretely, the cases $s=4, m=2$ and $s=5, m=2$ are studied in sequences A118647 and A120118 respectively.

Nevertheless, the cyclic version of this problem does not seem to have been studied anywhere in the literature. The following is an adaptation to our problem of the counting strategy briefly mentioned in the aforementioned references. It is based on counting the number of closed walks of length $k$ in certain graph.

Fix integers $m\geq 1$ and $s\geq 2$ with $m< s$.

\begin{Definition}
Let $V_{(s-1),m}$ be the set of all elements $x\in\{0,\dots,q-1\}^{s-1}$ of Hamming weight at most $m$.
We define the set $E_{(s-1),m}\subseteq V_{(s-1),m}\times V_{(s-1),m}$ as follows: $(x,y)\in E_{(s-1),m}$ if and only if
\begin{enumerate}
\item $x_2=y_1,\ x_3=y_2,\ \dots,\ x_{s-1}=y_{s-2}$ and
\item the weight of the string $(x_1,x_2,\dots,x_{s-1},y_{s-1})$ is at most $m$.
\end{enumerate}
\end{Definition}

 Now we have:

\begin{Theorem}\label{thm:graph}
Let $t\in\{0,\dots,q^k-1\}$ and denote its $q$-ary representation by $(t_0,t_1,\dots,t_{k-1})$. In addition for $i=0,\dots,s-1$, let $t_{k+i}:=t_i$ and $t^{[i,s]}:=(t_i,t_{i+1},\dots,t_{i+s-2})$. 

Then $\wqrel(t)\leq m$ if and only if, for all $j$ in $0,\dots,k-1$,
\begin{enumerate}
\item $t^{[j,s]}\in V_{(s-1),m}$ and
\item $(t^{[j,s]},t^{[j+1,s]})\in E_{(s-1),m}$
\end{enumerate}

\end{Theorem}

It is clear that the pair $(V_{(s-1),m},E_{(s-1),m})$ is a directed graph. In addition it has at most one edge connecting two vertices in a given direction, and it may contain loops (an edge may connect one vertex to itself).

In the following, for the sake of simplicity, a graph will mean a directed graph with the properties just mentioned.

\begin{Definition}
A walk of length $k$ in a graph $(V,E)$ is a sequence $(v_0, v_1,\dots, v_k)$ such that $v_j\in V$ for $j=0,\dots,k$ and $(v_{j-1},v_j)\in E$ for $j=1,\dots,k$. Here $v_j$ and $v_{j'}$ do not need to be different.
The vertex $v_0$ is the initial vertex of the walk and the vertex $v_k$ is the terminal vertex of the walk. The walk is closed if the initial and terminal vertices coincide, i.e., $v_0=v_k$.
\end{Definition}

\begin{Proposition}
There is a one to one correspondence between the set $\{t\in\{0,\dots,n-1\}: \wqrel(t)\leq m\}$ and the set of closed walks of length $k$ in the graph $(V_{(s-1),m},E_{(s-1),m})$.
\end{Proposition}

\begin{Proof}
Given an integer $t\in\{0,\dots,n-1\}$ with $(t_0,t_1,\dots,t_{k-1})$ as its $q$-ary representation, we associate the sequence $t^{[0,s]},t^{[1,s]},\dots,t^{[k,s]}$, where recall that $t^{[j,s]}=(t_j,t_{j+1},\dots,t_{j+s-2})\in \{0,\dots,q-1\}^{s-1}$. As usual the sums in the indices are modulo $k$. Then a direct consequence of Theorem~\ref{thm:graph} is that $\wrel(t)\leq m$ if and only if $(t^{[0,s]},\dots,t^{[k,s]})$ is a walk of length $k$. Moreover $t^{[0,s]}=t^{[k,s]}$, so it is a closed walk. It is clear that every closed walk of length $k$ corresponds to a unique integer.

\end{Proof}

Now it is a well known fact from graph theory that

\begin{Lemma}
Let $A\in\RR^{g\times g}$ be the adjacency matrix of the graph. The number of walks of length $k$ with initial vertex $v$ and terminal vertex $w$ is the $(v,w)$-th entry of the matrix $A^k$. In particular, the $(v,v)$-th entry of $A^k$ is the number of closed walks of length $k$ starting and ending in $v$.
\end{Lemma}

\begin{Corollary}
We have 

$$N_{k,s,m}=\Tr(A^k)$$

where $A$ is the adjacency matrix of the graph $(V_{(s-1),m}, E_{(s-1),m})$ and $\Tr$ denotes its trace, i.e., the sum of its diagonal elements.
\end{Corollary}

Note that the graph $(V_{(s-1),m},E_{(s-1),m})$ and therefore its adjacency matrix do not depend on $k$, and hence having fixed $s,m$ the matrix $A$ is completely determined. Furthermore a recurrence formula can be given for the successive powers of $A$, and hence for their traces.

\begin{Proposition}
Let $A\in\RR^{g\times g}$-matrix and $p(X)=\sum_{i=0}^g p_i X^i$ its characteristic polynomial. Then $$\sum_{i=0}^g p_i  \Tr(A^{i+j})=0.$$

\end{Proposition}
The proposition follows from Cayley-Hamilton theorem, which states that $p(A)=0$, i.e., $\sum_{i=0}^g p_i  A^{i}$ is the all-zero matrix. Multiplying by $A^j$ and using linearity of the trace yields the result. 

This leads to a recurrence formula for the numbers $N_{k,s,m}$. Since these are only defined for $k\geq s$, we introduce the following definition.
\begin{Definition}
Let $N'_{k,s,m}:=\Tr(A^k)$, where $A$ is the adjacency matrix of the graph $(V_{(s-1),m},E_{(s-1),m})$. 
\end{Definition}

\begin{Theorem}
For $k\geq s$, $N_{k,s,m}=N'_{k,s,m}$; and for all $k\geq g$ the numbers $N'_{k,s,m}$ satisfy the recurrence $$N'_{k,s,m}=-\sum_{j=1}^g p_{g-j} N'_{(k-j),s,m}$$  where $p(X)=\sum_{i=0}^g p_i X^i$ is the characteristic polynomial of the graph $(V_{(s-1),m},E_{(s-1),m})$.
\end{Theorem}

\section{Some concrete values for binary codes}\label{sec:concrete}
In this section we compute the parameters obtained for $q=2$ and certain specific choices for $s$ and $m$ in the construction from the previous section. First we consider the case $m=1$ and give general formulas, depending on $s$ and $k$, for the dimension and lower bounds for the minimum distance of the squares. For $s=3$ and the first few values of $k$, the resulting parameters are collected in Table~\ref{tab:s3m1}. Later we choose $m=2$ and $s=5$ and collect the results in Table~\ref{tab:s5m2}.

\subsection{Case $m=1$}

Remember that for each integer $k\geq s$, we are considering the cyclic code $C$ of length $n=2^k-1$, generated by the polynomial $g=(X^n-1)/\prod_ {i\in W_{k,3,1}} (X-\beta^i)$, where $W_{k,3,1}=\{t\in\{0,\dots,n-1\}: w_2^{(3)}(t)\leq 1\}$. 

We first determine the numbers  ${B}_{k,s,1}$, $\widehat{B}_{k,s,2}$ which, according to Theorem~\ref{thm:codesm}, yield bounds for the minimum distance of the codes $C$ and $C^{*2}$ respectively. Let $\ell=k\mod s$. The binary representation of $B_{k,s,1}$ starts with $\left\lfloor k/s \right\rfloor$ blocks of the form $100...0$ (one $1$ and $s-1$ zeros). The remaining $\ell$ bits need to be all zero because otherwise the block of the last $s-1\geq \ell$ bits, together with the first one, would create a sequence of $s$ consecutive positions with weight at least $2$. 

Therefore a recurrence is given by the formula

$$
\recursion{B_{s,s,1}}{2^{s-1}}{B_{k,s,1}}{2B_{(k-1),s,1}+2^{s-1}}{k=0\mod s}{2B_{(k-1),s,1}}{k\neq 0\mod s}{k>s}
$$

If we now write $d_k= 2^k-1-B_{k,s,1}$ (which is the bound for $d(C)$ promised by Theorem~\ref{thm:codesmz}), then $d_k$ satisfies
$$
\recursion{d_s}{2^{s-1}-1}{d_k}{2d_{k-1}-2^{s-1}+1}{ k=0\mod s}{2d_{k-1}+1}{k\neq 0 \mod s}{k>s}
$$

In the case of the numbers $\widehat{B}_{k,s,2}$, observe that, since $m=1$ and hence $2m\left\lfloor\frac{k}{s}\right\rfloor=2\left\lfloor\frac{mk}{s}\right\rfloor $, the second summand of the expression in Lemma~\ref{lem: bsm} is always 0.

Therefore, the binary representation of $\widehat{B}_{k,s,2}$ consists of $\left\lfloor k/s \right\rfloor$ blocks of the form $110...0$ (two $1$'s and $s-2$ 0's) followed by $\ell$ 0's.
Hence $\widehat{B}_{k,s,2}$ satisfies the recurrence

$$
\recursion{\widehat{B}_{s,s,2}}{2^{s-1}+2^{s-2}}{\widehat{B}_{k,s,2}}{2B_{(k-1),s,2}+3\cdot 2^{s-2}}{k=0\mod s}{2B_{(k-1),s,2}}{k\neq 0 \mod s}{k>s}
$$

Moreover, the numbers $\widehat{d}_k= 2^k-1-\widehat{B}_{k,s,2}$ satisfy

$$
\recursion{\widehat{d}_s}{2^{s-2}-1}{\widehat{d}_k}{2\widehat{d}_{k-1}-3\cdot 2^{s-2}+1}{k=0\mod s}{2\widehat{d}_{k-1}+1}{k\neq 0 \mod s}{k>s}
$$

We now determine the numbers $N'_{k,s,1}$ which for $k\geq s$ yield the dimension of the code (the size of $I$). For this we use the graph $(V_{(s-1),1},E_{(s-1),1})$.

The vertex set $V_{(s-1),1}$ consists of the all-zero vector and all the unit vectors in $\{0,1\}^s$.
 The graph $(V_{(s-1),1},E_{(s-1),1})$ is

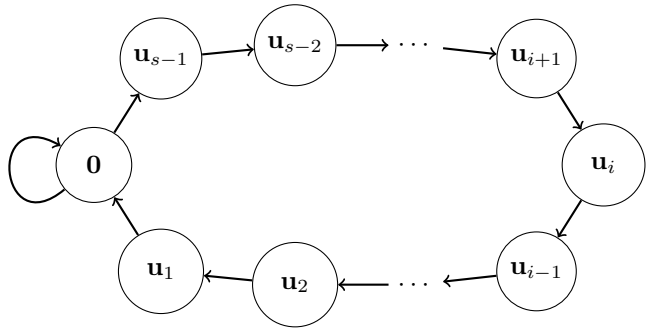
\begin{center}
\begin{tikzpicture}
\node (a) at (25bp,59bp) [draw,circle,inner sep=7pt] {$\bf{0}$};
\node (b) at (50bp,99bp) [draw,circle,] {$\mathbf{u}_{s-1}$};
\node (c) at (100bp,104bp) [draw,circle,] {$\mathbf{u}_{s-2}$};
\node (d) at (145bp,104bp){$\cdots$};
\node (l) at (190bp,99bp)  [draw,circle,] {$\mathbf{u}_{i+1}$};
\node (m) at (215bp,59bp)  [draw,circle,inner sep=7pt] {$\mathbf{u}_{i}$};
\node (n) at (190bp,19bp) [draw,circle] {$\mathbf{u}_{i-1}$};
\node (x) at (145bp,14bp){$\cdots$};
\node (y) at (100bp,14bp) [draw,circle,inner sep=7pt] {$\mathbf{u}_2$};
\node (z) at (50bp,19bp) [draw,circle,inner sep=7pt] {$\mathbf{u}_1$};

\path[]
(a) edge[->,thick] (b)
(b) edge[->,thick] (c)
(c) edge[->,thick] (d)
(d) edge[->,thick] (l)
(l) edge[->,thick] (m)
(m) edge[->,thick] (n)
(n) edge[->,thick] (x)
(x) edge[->,thick] (y)
(y) edge[->,thick] (z)
(z) edge[->,thick] (a)
(a) edge[->,thick, loop below, in=150, out=220, looseness=5] (a);

\end{tikzpicture}
\end{center}

Indeed observe that even though $\mathbf{u}_1$ can be glued with $\mathbf{u}_{s-1}$, the resulting vector $10...01$ would have weight $2$ and hence $(\mathbf{u}_1,\mathbf{u}_{s-1})\notin E_{(s-1),1}$. 

The adjacency matrix $A$ of the graph is of the following form

$$
\begin{pmatrix}
1&1&0&0&\dots&0&0\\
0&0&1&0&\dots&0&0\\
0&0&0&1&\dots&0&0\\
\vdots&\vdots&\vdots&\vdots&\ddots&\vdots&\vdots\\

0&0&0&0&\dots&1&0\\
0&0&0&0&\dots&0&1\\
1&0&0&0&\dots&0&0\\
\end{pmatrix}
$$

It is not difficult to verify that the characteristic polynomial of $A$ is $X^s-X^{s-1}-1$.

Hence we have the recurrence $$N'_{k,s,1}=N'_{(k-1),s,1}+N'_{(k-s),s,1},$$ for $k\geq s$. It remains to compute the values $N'_{k,s,1}=\Tr(A^k)$ for $0\leq k\leq s-1$. Observe that $\Tr(A^0)=s$ and $Tr(A)=1$ can be observed directly. For the remaining values, one could compute the matrix $A^k$, but it is just easier to remember that $\Tr(A^k)$ is the number of closed walks of length k in the graph. Clearly, for $k< s$, the only closed walk of length $k$ is the walk $(\mathbf{0},\mathbf{0},\dots,\mathbf{0})$, as any walk involving any other vertex will take at least $k$ steps to return to the origin. Hence, $N'_{k,s,1}=1$ for $1\leq k\leq s-1$.

The observations in this section are collected in the following theorem.

\begin{Theorem}\label{thm:s1}
Let $k\geq s\geq 3$. Let $C$ be the cyclic code generated by the polynomial $g=(X^n-1)/f(X)$, where $f=\prod_{i\in W_{k,s,1}} (X-\beta^i)$.
Then
\begin{itemize}
\item $\dim C=N'_{k,s,1}$, where $N'_{k,s,1}$ is given by the recurrence 
 $N'_{0,s,1}=s$, $N'_{k,s,1}=1$ for $1\leq k\leq s-1$, and

$$N'_{k,s,1}=N'_{(k-1),s,1}+N'_{(k-s),s,1}, \textrm{ for } k\geq s.$$  

\item $d(C)\geq d_k,$ where $d_k$ is given by the recurrence
$$
\recursion{d_s}{2^{s-1}-1}{d_k}{2d_{k-1}-2^{s-1}+1}{k=0\mod s}{2d_{k-1}+1}{k\neq 0 \mod s}{k>s}
$$

\item $d(C^{*2})\geq \widehat{d}_k,$ where $\widehat{d}_k$ is given by the recurrence 

$$
\recursion{\widehat{d}_s}{2^{s-2}-1}{\widehat{d}_k}{2\widehat{d}_{k-1}-3\cdot 2^{s-2}+1}{k=0\mod s}{2\widehat{d}_{k-1}+1}{k\neq 0 \mod s}{k>s}
$$

\end{itemize}

\end{Theorem}

The explicit parameters obtained for the first few values of $k$ in the case $s=3$ are collected in Table \ref{tab:s3m1}. Here $\dim C$ and the bounds for $d(C)$ and $d(C^{*2})$ follow from the explicit formulas above, while the values for $\dim C^{*2}$ have been obtained by direct computation.

\begin{table}[h]
\begin{center}
\begin{tabular}{|c|c|c|c|c|c|c|}
\hline
$k$ & $n$ & $\dim C$ & $d(C)\geq $& $\dim C^{*2}$& $d(C^{*2})\geq$\\ \hline
$3$ & $7$ & $4$ & $3$ & $7$ & $1$\\
$4$ & $15$ & $5$ & $7$ & $11$& $3$\\
$5$ & $31$ & $6$ & $15$ & $16$& $7$\\
$6$ & $63$ & $10$ & $27$&$37$& $9$\\  
$7$ & $127$ & $15$ & $55$&$71$& $19$\\ 
$8$ & $255$&$21$&$111$&$123$&$39$\\
$9$ & $511$&$31$&$219$&$232$&$73$\\ 
$10$ & $1023$&$46$&$439$&$441$& $147$ \\ 
$11$ & $2047$&$67$&$879$& $804$&$295$\\ 
$12$ & $4095$&$98$&$1755$&$1475$& $585$\\ 
\hline
\end{tabular}

\bigskip
\begin{tabular}{|c|c|c|}
\hline
$k$ & $n$ & Observations\\ \hline
$3$ & $7$ & Both $C$ and $C^{*2}$ optimal \\
$4$ & $15$ & Both $C$ and $C^{*2}$ optimal\\
$5$ & $31$ & $C$ optimal, $C^{*2}$ not\\
$6$ & $63$ & $C$ best known, $C^{*2}$ not\\  
$7$ & $127$ & Both $C$ and $C^{*2}$ best known\\ 
$8$ & $255$& Both $C$ and $C^{*2}$ best known\\
$9$ & $511$& \\ 
$10$ & $1023$&\\ 
$11$ & $2047$& \\ 
$12$ & $4095$&\\

\hline
\end{tabular}

\end{center}
\caption{Case $m=1$, $s=3$.}
\label{tab:s3m1}
\end{table}

Moreover, the parameters obtained for both $C$ and $C^{*2}$ are compared with the code tables from \cite{Gra, Min} which collect lower and upper bounds for the largest possible minimum distance of a (in this case, binary) linear code of a given length and dimension. In the table below the observation ``$C$ (resp. $C^{*2}$) best known"  means that, according to \cite{Gra, Min}, no binary code of length $n$ is known that has the same dimension of $C$ (resp. $C^{*2}$) and larger minimum distance. Furthermore, ``optimal" means that no code with the same length and dimension and strictly larger minimum distance can exist.

\begin{Remark}\label{rem:exactd}
If $3$ divides $k$, and $C$ is defined as in this section, then there is a word in $C^{*2}$ of weight $n/7$. Therefore for $k=3,6,9,12$, the corresponding entry in the table is actually the true value of $d(C^{*2})$. 
\end{Remark}
Indeed if $3$ divides $k$, then  $n/7=(2^k-1)/(2^3-1)=1+8+64+\cdots+8^{k/3-1}$. The polynomial $f=(X^n-1)/(X^{n/7}-1)$ is such that $f(\beta^i)=0$ if $7$ does not divide $i$, and $f(\beta^i)=1$ if $7$ divides $i$. Note however that $f=1+X^{n/7}+\dots+X^{6n/7}$ and that 
$n-jn/7=(7-j)+8(7-j)+64(7-j)+\cdots+8^{k/3-1}(7-j)$, $j=1,\dots,6$. From here one can easily find the binary representation of $n-jn/7$ and argue that $jn/7\in -(I+I)$ for $j=1,\dots,6$; obviously $0\in -(I+I)$ too. Hence $f\in{\cal P}(-(I+I))$ and its vector of evaluations is an element in $\B(-(I+I))$ defined over $\FF_2$, thus it is a word of $C^{*2}$. By the observation above it has weight $n/7$.

\begin{Remark}[Asymptotics]
In fact, $d(C^{*2})\geq n/7$ in all cases and $d(C)$, $d(C^{*2})=\Theta(n)$. On the other hand it is easy to see that $\dim C= \Theta(n^{\log_2\gamma})=\Theta(n^{0.551...})$, where $\gamma=1.465...$ is the only real root of $X^3-X^2-1$, an eigenvalue of the matrix $A$.
\end{Remark}

\subsection{Case $s=5$, $m=2$}
In order to obtain codes with larger dimension (for the same length), one needs to increase the value of $m$. On the other hand, fixing a value of $m$, the largest dimensions are obtained when $s$ is as small as possible, and since we are operating under the restriction $m\leq \frac{s-1}{2}$, this suggests to use $s=2m+1$. In this section the case $s=5$, $m=2$ is analysed.

The first $5\lfloor k/5 \rfloor$ bits of the binary representations of the numbers $B_{k,5,2}$ consist of $\lfloor k/5 \rfloor$ repetitions of the block $11000$. The remaining bits must satisfy that the three last bits need to be $0$ because of the restricted weight (cyclic) constraint and the fact that the two first bits of $B_{k,5,2}$ are 1. Therefore, these remaining bits are respectively $0, 00, 000, 1000$ for $k=1, 2, 3, 4\mod 5$.
Hence we have the recurrence 
$$\recursion{B_{5,5,2}}{24}{B_{k,5,2}}{2B_{(k-1),5,2}+8}{k=0,4\mod 5}{2B_{(k-1),5,2}}{k=1, 2, 3\mod 5}{k>5}
$$
As for $\widehat{B}_{k,5,4}$, note that the first $5\lfloor k/5 \rfloor$ bits of their binary representations are $11110$. The conditions on the restricted weight imply that the last bit must be $0$ of each of these numbers. Finally, by the definition the binary weight of 
$\widehat{B}_{k,5,4}$ is at most $2\lfloor 2k/5\rfloor$, which equals $4\lfloor k/5\rfloor$ if $k=0,1,2 \mod 5$ and $4\lfloor k/5\rfloor+2$ if $k=3,4\mod 5$. Note that the first $5\lfloor k/5 \rfloor$ bits of $\widehat{B}_{k,5,4}$ already have weight $4\lfloor k/5\rfloor$.
Hence the remaining bits are respectively $0, 00, 110, 1100$ for $k=1, 2, 3, 4\mod 5$, and we have the recurrence

$$\recursion{\widehat{B}_{5,5,4}}{30}{\widehat{B}_{k,5,4}}{2\widehat{B}_{(k-1),5,4}+6}{k=0,3\mod 5}{2\widehat{B}_{(k-1),5,4}}{k=1,2,4\mod 5}{k>5}$$

We analyse the numbers $N'_{k,5,2}$. The set $V_{4,2}$ consists of the 11 vectors $0000$, $0001$, $0010$, $0011$, $0100$, $0101$, $0110$, $1000$, $1001$, $1010$, $1100$. The characteristic polynomial of the graph $(V_{4,2},E_{4,2})$ is $X^{11}-X^{10}-X^8-2X^6+X^3+X$ and therefore the recurrence 

$$N'_{k,5,2}=N'_{(k-1),5,2}+N'_{(k-3),5,2}+2N'_{(k-5),5,2}-N'_{(k-8),5,2}-N'_{(k-10),5,2}$$

 holds for $k\geq 11$.

Direct computation yields that the values of $N'_{k,5,2}$ for $k=1,2,\dots,10$ are $1,1,4,5,16,22,29,45,76,126$ respectively.

\begin{Theorem}
Let $k\geq 5$. Let $C$ be the cyclic code generated by the polynomial $g=(X^n-1)/f(X)$, where $f=\prod_{i\in W_{k,5,2}} (X-\beta^i)$.
Then

\begin{itemize}
\item $\dim C=N'_{k,5,3}$ where, for $k=1,2,\dots,10$, respectively  $N'_{k,5,2}=1,1,4,5,16,22,29,45,76,126$ and, for $k\geq 11$,
$$
N'_{k,5,2}=N'_{(k-1),5,2}+N'_{(k-3),5,2}+2N'_{(k-5),5,2}-N'_{(k-8),5,2}-N'_{(k-10),5,2}.$$


\item $d(C)\geq d_k,$ where $d_k$ is given by the recurrence 
$$\recursion{d_5}{7}{d_k}{2d_{k-1}-7}{k=0,4\mod 5}{2d_{k-1}+1}{k=1,2,3\mod 5}{k>5} $$

\item $d(C^{*2})\geq \widehat{d}_k,$ where $\widehat{d}_k$ is given by the recurrence 

$$\recursion{\widehat{d}_5}{1}{\widehat{d}_k}{2\widehat{d}_{k-1}-5}{k=0,3\mod 5}{2\widehat{d}_{k-1}+1}{k=1,2,4\mod 5}{k>5}$$

\end{itemize}

\end{Theorem}

Concretely, for the first few values of $k$, we obtain the parameters collected by Table~\ref{tab:s5m2}. The same comments about the ``Observations" column apply as in Table~\ref{tab:s3m1}.

\begin{table}[h]
\begin{center}
\begin{tabular}{|c|c|c|c|c|c|}
\hline
$k$& $n$  & $\dim C$ & $d(C)\geq $& $\dim C^{*2}$ & $ d(C^{*2})\geq$\\ 
\hline
$5$&$31$&$16$&$7$& $31$& $1$\\
$6$&$63$& $22$&$15$& $57$ &$3$\\
$7$&$127$ &$29$&$31$ &$99$ &$7$\\  
$8$&$255$&$45$& $63$& $223$ &$9$\\ 
$9$&$511$&$76$& $119$ & $430$ & $19$\\ 
$10$&$1023$&$126$& $231$ &$863$ &$33$ \\ 
$11$&$2047$&$210$& $463$ & $1695$ &$67$\\ 
$12$&$4095$&$338$ &$927$& $3293$ & $135$ \\

\hline
\end{tabular}

\bigskip

\begin{tabular}{|c|c|c|}
\hline
$k$& $n$  & Observations\\ 
\hline
$5$&$31$&$C$ best known, $C^{*2}$ optimal\\
$6$&$63$& $C^{*2}$ optimal\\
$7$&$127$ &\\  
$8$&$255$& $C^{*2}$ best known\\ 
$9$&$511$& $C^{*2}$ best known\\ 
$10$&$1023$& $C^{*2}$ best known \\ 
$11$&$2047$&\\ 
$12$&$4095$& \\

\hline
\end{tabular}

\end{center}
\caption{Case $m=2$, $s=5$.}
\label{tab:s5m2}
\end{table}

Similarly as in Remark~\ref{rem:exactd} one can argue

\begin{Remark}
If $5$ divides $k$, then there is a word in $C^{*2}$ of weight $n/31$. Therefore for $k=5,10$, the corresponding entry in the table is actually the true value of $d(C^{*2})$. 
\end{Remark}

\begin{Remark}[Asymptotics]
We have that $d(C), d(C^{*2})=\Theta(n)$, where in fact $d(C^{*2})\geq n/31$ for all $n$. Moreover $\dim C=\Theta(n^{\log_2 \gamma})=\Theta(n^{0.697...})$ where $\gamma=1.622...$ is the only real eigenvalue $\gamma>1$ of the matrix $A$. 
\end{Remark}

Finally, note that by Theorem~\ref{thm:codesmz}, for every entry of Tables~\ref{tab:s3m1} and \ref{tab:s5m2} and its corresponding code $C$, another cyclic code $C'$ can be found with the same length, with $\dim C'=\dim C-1$ and such that the lower bounds for $d(C')$ and $d((C')^{*2})$ are one unit more than in the table.
Furthermore, Proposition~\ref{prop:puncshor} guarantees that if $a,b$ are integers with as long as $a+b<n$ and $b<d(C)$, then by shortening and puncturing we can find a (non-necessarily cyclic) code $D$ with length $n-a-b$, $\dim D\geq \dim C-a$,  $d(D)\geq d(C)-b$ and $d(D^{*2})\geq d(C^{*2})-b$. A similar remark holds by replacing $C$ by the $C'$ mentioned some lines above. 

\subsection{Remarks and comparisons}

In some cases in Tables~\ref{tab:s3m1} and \ref{tab:s5m2}, both $C$ and $C^{*2}$ are optimal in the sense that both $d(C)$ is the largest possible for a code of length $n$ and dimension $\dim C$, and $d(C^{*2})$ is the largest possible for a code of length $n$ and dimension $\dim C^{*2}$. In other cases, both $d(C)$ and $d(C^{*2})$ match the largest values which are known to be attainable according to the tables of binary codes in~\cite{Gra, Min}. It should be remarked that~\cite{Gra, Min} only contains information about binary linear codes up to certain length (which is 512 in~\cite{Gra} and 1024 under certain restrictions for the dimension in the case of~\cite{Min}) and hence the parameteres of some of the longer codes obtained here cannot be measured against these tables.

A natural question is whether it also holds that $d(C^{*2})$ is the largest possible given $(n, \dim C)$, since this would be desirable for the applications mentioned in the introduction. However, this cannot be established from the optimality of $C$ and $C^{*2}$ only, since it is conceivable that there exists another code $E$ of length $n$ such that $\dim E=\dim C$ and $\dim E^{*2}<\dim C^{*2}$; in such a case it would be possible that $d(E^{*2})>d(C^{*2})$. It is therefore unclear whether the codes in the table do achieve the largest possible value for $d(C^{*2})$ given $(n, \dim C)$ and it is left as an open question. 

We now compare the codes in the table with other families of linear codes. First, for given values of $k,m$, note that the construction that we are considering contains a code obtained from Reed-Muller code $RM(m,k)$ by puncturing one position. This is a consequence of the fact that any integer $t$ satisfying $w_2(t)\leq m$ also satisfies $w_2^{(s)}(t)\leq m$, together with the observations from Proposition~\ref{prop:grm}. Consequently the dimension of the corresponding entry $C$ in the table will be at least that of $RM(m,k)_{\bullet}$ (which denotes puncturing $RM(m,k)$ in one position), while $d(C^{*2})$ will be at most $d(RM(m,k)_{\bullet}^{*2})= 2^{k-2m}-1$. For small values of $k$ (concretely $k\leq 5$ in the first table and $k\leq 7$ in the second), the code from the table is actually the same as the punctured Reed-Muller code, since in those cases there are no other integers with $w_2^{(s)}(t)\leq m$. We also compare the codes in the tables with Reed Muller codes of the form $RM(m',k)$ where $m'\neq m$. One can see then that in all cases at least one of the two parameters $\dim C$, $d(C^{*2})$ of a code in the table is better than that of a punctured Reed-Muller code $RM(m',k)_{\bullet}$ of the same length. In some cases both parameters are better. For example, the code $RM(3,12)_{\bullet}$ of length 2047, satisfies $\dim RM(3,12)_{\bullet}=299$ and $d(RM(3,12)_{\bullet}^{*2})=63$. These parameters are both worse than the entry of the same length in  Table~\ref{tab:s5m2}.

 In some cases we can analyse how the parameters of the codes in the tables compare to binary codes from Corollary~\ref{cor:RSconcatenate} and the second part of Proposition~\ref{prop:initial}. Proposition~\ref{prop:initial} yields codes of length $n$ with $d(C^{*2})\cdot \dim C=n$. All the entries in the tables, except the first two entries in Table~\ref{tab:s3m1} and the first entry in Table~\ref{tab:s5m2} satisfy that $d(C^{*2})\cdot\dim C>n$. Moreover, one can apply the first part of Proposition~\ref{prop:initial} to the codes in the tables and all the resulting codes will likewise satisfy $d(C^{*2})\cdot\dim C>n$. For parameters such as the ones mentioned in the introduction, this means we can find shorter codes for a specified bound on $\dim C$ and $d(C^{*2})$, as soon as these bounds are large enough. For example, if one needs a code $C$ with $\dim C\geq 200$ and $d(C^{*2})\geq 60$, Proposition~\ref{prop:initial} would require its length to be at least 12000, while Table~\ref{tab:s5m2} shows there is a cyclic code with length 2047 satisfying such properties. Finally we consider the binary case of Corollary~\ref{cor:RSconcatenate}. We restrict ourselves to the case $s\leq 3$ of the corollary, as otherwise the resulting codes are very long ($n>20000$). In that case, it turns out that for all selections of the values $s,m$ in the corollary, we have  $d(C^{*2})\cdot \dim C\leq n$ for the resulting code (and similar considerations as before apply), except when $s=3$ and $m=6$ or $7$. In these two last cases, we have $n=3584$, and respectively $(\dim C=49, d(C^{*2})=74)$ and $(\dim C=56, d(C^{*2})=65)$. Then we can see for example that the code of length 2047 in Table~\ref{tab:s3m1} is shorter and has much larger dimension and minimum distance of the square than both of the aforementioned codes.
In conclusion, at least for the cases analysed in this section, the constructions in this paper complement the ones we can obtain from Reed Muller codes and compare favourably to other constructions in this range.

\section{Acknowledgments}
The author would like to thank Ren\'e B\o dker Christensen, Jaron Skovsted Gundersen and Diego Ruano for helpful discussions, and the anonymous reviewers of IEEE Transactions on Information Theory for their comments, which have improved the quality of the paper. The author is especially grateful to one of the reviewers for identifying an omission regarding Theorem~\ref{thm:squarescyclic} and suggesting the idea for its proof.

\begin{appendix}

\section{Proof of Proposition~\ref{prop:srw}}

\begin{Proof}
The case $s=k$ is Lemma~\ref{lem:weights}, so we assume $s\leq k-1$, which simplifies the notation.

Let $v:=t+u \mod n$, and remember $n=q^k-1$. Since the restricted weights are invariant of cyclotomic cosets, we can assume without loss of generality that the maximum in the definition of $\wqrel(v)$ is attained for the set of $s$ least significant digits. That is, if we consider $v$ as an integer in $\{0,\dots,q^k-2\}$ and write $v=v'+v''q^s$, where $0\leq v'\leq q^s-1$ and $0\leq v''\leq q^{k-s}-1$, then we are assuming $\wqrel(v)=\wq(v')$. 
We also write $t=t'+t''q^s$, $u=u'+u''q^s$, where $0\leq t',u'\leq q^s-1$ and $0\leq t'',u''\leq q^{k-s}-1$. Note that $\wq(t')\leq \wqrel(t)$ and $\wq(u')\leq \wqrel(u)$.

We now need to split the proof in different cases, according to whether $t+u$ (summed over the integers) is smaller than, equal to or larger than $q^k-1$.\\

Case 1. $t+u\leq q^k-2$.

In this case $v=t+u$ and therefore $t'+u'=v'+\epsilon q^s$ where $\epsilon=0$ or $1$. Now clearly $\wqrel(v)=\wq(v')\leq \wq(v')+\epsilon=\wq(v'+\epsilon q^s)=\wq(t'+u')\leq \wq(t')+\wq(u')\leq \wqrel(t)+\wqrel(u)$, where the inequality $\wq(t'+u')\leq \wq(t')+\wq(u')$ comes from Lemma~\ref{lem:weights}.\\

Case 2. $t+u=q^k-1$.

In this case $v=0$ and the statement follows trivially since $\wqrel(v)=0$ and all weights are non-negative.\\

Case 3. $t+u\geq q^k$.

This case is more involved. Note $v=t+u+1-q^k$. Then the $q$-ary representation of $v$ is obtained by computing the one for $t+u+1$ and then erasing the $1$ in the position corresponding to $q^k$. It is easy to see then that $t'+u'+1=v'+\epsilon q^s$, where $\epsilon=0$ or $1$.
 We need to further split the proof in these two cases.\\

Case 3a. $t+u\geq q^k$ and $t'+u'+1\geq q^s$.

In this case (since also $t'+u'+1\leq 2q^s-1$) it holds that $t'+u'+1=v'+ q^s$.
Then $\wqrel(v)+1=\wq(v')+1=\wq(v'+\epsilon q^s)=\wq(t'+u'+1)\leq \wq(t')+\wq(u')+1\leq \wqrel(t)+\wqrel(u)+1$ and hence $\wqrel(v)\leq \wqrel(t)+\wqrel(u)$.\\

Case 3b. $t+u\geq q^k$ and $t'+u'+1\leq q^s-1$.

In this case it holds that $t'+u'+1=v'$ and we can only show the inequality  $\wq(v')\leq \wq(t')+\wq(u')+1$. In fact, the inequality is tight, i.e., there are cases in which $\wq(v')=\wq(t')+\wq(u')+1$. 

In order to show the theorem, we need to argue the following:\\

\begin{claim}
Under the restrictions of case 3b., it holds that $\wqrel(t)+\wqrel(u)\geq \wq(t')+\wq(u')+1$.
\end{claim}

Once we prove this claim, the proof is finished, since in that case $\wqrel(v)=\wq(v')\leq \wq(t')+\wq(u')+1\leq \wqrel(t)+\wqrel(u)$.

\begin{Proofc}
Clearly $\wqrel(t)\geq \wq(t')$ and $\wqrel(u)\geq \wq(u')$. So we need to rule out either $\wqrel(t)= \wq(t')$ or $\wqrel(u)=\wq(u')$. Write $t=\sum_{i=0}^{k-1} t_iq^i$, $u=\sum_{i=0}^{k-1} u_iq^i$. Then proving the claim amounts to showing the existence of $j\in\{1,\dots,k+1\}$ such that either $\sum_{i=0}^{s-1} t_{j+i}>\sum_{i=0}^{s-1} t_i(=\wq(t'))$, or $\sum_{i=0}^{s-1} u_{j+i}>\sum_{i=0}^{s-1} u_i(=\wq(u'))$, where the sums $j+i$ are modulo $k$.

Suppose towards a contradiction, that this is not true, and hence $\wqrel(t)+\wqrel(u)= \wq(t')+\wq(u')$. We now make the following claim.

\begin{claim}
Under the restrictions of case 3b. and assuming $\wqrel(t)+\wqrel(u)=\wq(t')+\wq(u')$, we have $t_{k-j}+u_{k-j}=t_{s-j}+u_{s-j}=q-1$ for all $j\in\{1,\dots,s\}$.
\end{claim}

Assuming claim 2, we quickly arrive at a contradiction, since in fact in that case $t'+u'=(q-1)(1+q+\dots+q^{s-1})=q^s-1$ but we are assuming $t'+u'+1\leq q^s-1$. This shows claim 1.
Hence we are left to prove claim 2.

\begin{Proofc}
We argue by induction on $j$.
 
For the case $j=1$, note that the condition $t'+u'+1\leq q^s-1$ clearly implies that $t_{s-1}+u_{s-1}\leq q-1$. On the other hand since $t=t_{k-1}q^{k-1}+\tilde{t}$,  $u=u_{k-1}q^{k-1}+\tilde{u}$ with $\tilde{t}, \tilde{u}\leq q^{k-1}-1$ the condition $t+u\geq q^k$ implies that $t_{k-1}+u_{k-1}\geq q-1$. However, if $t_{s-1}+u_{s-1}< t_{k-1}+u_{k-1}$, then $\sum_{i=0}^{s-1} t_i+\sum_{i=0}^{s-1} u_i<\sum_{i=0}^{s-1} t_{k-1+i}+\sum_{i=0}^{s-1} u_{k-1+i}$, and we reach a contradiction. So the only possibility is $t_{s-1}+u_{s-1}=t_{k-1}+u_{k-1}=q-1$.

Now, assume $t_{k-j}+u_{k-j}=t_{s-j}+u_{s-j}=q-1$ is true for all $j<j_*$. Thus we have $q^s-1\geq t'+u'\geq(q-1)(q^{s-1}+\dots+q^{s-j_*+1})+(t_{s-j_*}+u_{s-j_*})q^{s-j_*}$. Then it is easy to see that this implies $t_{s-j_*}+u_{s-j_*}\leq q-1$. On the other hand $q^k\leq t+u<(q-1)(q^{k-1}+\dots+q^{k-j_*+1})+(t_{k-j_*}+u_{k-j_*})q^{k-j_*}+2q^{k-j_*}=q^k-q^{k-j_*+1}+(t_{k-j_*}+u_{k-j_*}+2)q^{k-j_*}$. This implies $t_{k-j_*}+u_{k-j_*}> q-2$, hence $t_{k-j_*}+u_{k-j_*}\geq q-1$.
Finally by the assumption $\wqrel(t)+\wqrel(u)= \wq(t')+\wq(u')$, we have  $\sum_{i=0}^{s-1} t_i+\sum_{i=0}^{s-1} u_i\geq \sum_{i=0}^{s-1} t_{k-j_*+i}+\sum_{i=0}^{s-1} u_{k-j_*+i}$. However, taking into account the induction assumption $t_{k-j}+u_{k-j}=t_{s-j}+u_{s-j}=q-1$ for $j<j_*$ and after removing terms that appear in both sides, we have $t_{s-j_*}+u_{s-j_*}\geq t_{k-j_*}+u_{k-j_*}$. Then necessarily $t_{s-j_*}+u_{s-j_*}= t_{k-j_*}+u_{k-j_*}=q-1$ and we have completed the induction and shown claim 2.

\end{Proofc}
\end{Proofc}

\end{Proof}

\end{appendix}


\begin{thebibliography}{10}

\bibitem{Bie02}
J{\"{u}}rgen Bierbrauer.
\newblock The theory of cyclic codes and a generalization to additive codes.
\newblock {\em Des. Codes Cryptography}, 25(2):189--206, 2002.

\bibitem{Cas15}
Ignacio Cascudo.
\newblock Powers of codes and applications to cryptography.
\newblock In {\em 2015 {IEEE} Information Theory Workshop, {ITW} 2015,
  Jerusalem, Israel, April 26 - May 1, 2015}, pages 1--5, 2015.

\bibitem{CCCX09}
Ignacio Cascudo, Hao Chen, Ronald Cramer, and Chaoping Xing.
\newblock Asymptotically {G}ood {I}deal {L}inear {S}ecret {S}haring with
  {S}trong {M}ultiplication over \emph{{A}ny} {F}ixed {F}inite {F}ield.
\newblock In {\em Advances in Cryptology - {CRYPTO} 2009, 29th Annual
  International Cryptology Conference, Santa Barbara, CA, USA, August 16-20,
  2009. Proceedings}, pages 466--486, 2009.

\bibitem{CCMZ15}
Ignacio Cascudo, Ronald Cramer, Diego Mirandola, and Gilles Z{\'{e}}mor.
\newblock Squares of {R}andom {L}inear {C}odes.
\newblock {\em {IEEE} Trans. Information Theory}, 61(3):1159--1173, 2015.

\bibitem{CCX14}
Ignacio Cascudo, Ronald Cramer, and Chaoping Xing.
\newblock Torsion {L}imits and {R}iemann-{R}och {S}ystems for {F}unction
  {F}ields and {A}pplications.
\newblock {\em {IEEE} Trans. Information Theory}, 60(7):3871--3888, 2014.

\bibitem{CDD16}
Ignacio Cascudo, Ivan Damg{\aa}rd, Bernardo David, Nico D{\"{o}}ttling, and
  Jesper~Buus Nielsen.
\newblock Rate-1, {L}inear {T}ime and {A}dditively {H}omomorphic {UC}
  {C}ommitments.
\newblock In {\em Advances in Cryptology - {CRYPTO} 2016 - 36th Annual
  International Cryptology Conference, Santa Barbara, CA, USA, August 14-18,
  2016, Proceedings, Part {III}}, pages 179--207, 2016.

\bibitem{CDD15}
Ignacio Cascudo, Ivan Damg{\aa}rd, Bernardo~Machado David, Irene Giacomelli,
  Jesper~Buus Nielsen, and Roberto Trifiletti.
\newblock Additively {H}omomorphic {UC} {C}ommitments with {O}ptimal
  {A}mortized {O}verhead.
\newblock In {\em Public-Key Cryptography - {PKC} 2015 - 18th {IACR}
  International Conference on Practice and Theory in Public-Key Cryptography,
  Gaithersburg, MD, USA, March 30 - April 1, 2015, Proceedings}, pages
  495--515, 2015.

\bibitem{CC06}
Hao Chen and Ronald Cramer.
\newblock Algebraic geometric secret sharing schemes and secure multi-party
  computations over small fields.
\newblock In {\em Advances in Cryptology - {CRYPTO} 2006, 26th Annual
  International Cryptology Conference, Santa Barbara, California, USA, August
  20-24, 2006, Proceedings}, pages 521--536, 2006.

\bibitem{COT17}
Alain Couvreur, Ayoub Otmani, and Jean{-}Pierre Tillich.
\newblock Polynomial time attack on wild mceliece over quadratic extensions.
\newblock {\em {IEEE} Trans. Information Theory}, 63(1):404--427, 2017.

\bibitem{CDM00}
Ronald Cramer, Ivan Damg{\aa}rd, and Ueli~M. Maurer.
\newblock General secure multi-party computation from any linear secret-sharing
  scheme.
\newblock In {\em Advances in Cryptology - {EUROCRYPT} 2000, International
  Conference on the Theory and Application of Cryptographic Techniques, Bruges,
  Belgium, May 14-18, 2000, Proceeding}, pages 316--334, 2000.

\bibitem{CDN15}
Ronald Cramer, Ivan Damg{\aa}rd, and Jesper~Buus Nielsen.
\newblock {\em Secure Multiparty Computation and Secret Sharing}.
\newblock Cambridge University Press, 2015.

\bibitem{DDGN14}
Ivan Damg{\aa}rd, Bernardo~Machado David, Irene Giacomelli, and Jesper~Buus
  Nielsen.
\newblock Compact {VSS} and efficient homomorphic {UC} commitments.
\newblock In {\em Advances in Cryptology - {ASIACRYPT} 2014 - 20th
  International Conference on the Theory and Application of Cryptology and
  Information Security, Kaoshiung, Taiwan, R.O.C., December 7-11, 2014,
  Proceedings, Part {II}}, pages 213--232, 2014.

\bibitem{DNNR16}
Ivan Damg{\aa}rd, Jesper~Buus Nielsen, Michael Nielsen, and Samuel Ranellucci.
\newblock The tinytable protocol for 2-party secure computation, or:
  Gate-scrambling revisited.
\newblock In {\em Advances in Cryptology - {CRYPTO} 2017 - 37th Annual
  International Cryptology Conference, Santa Barbara, CA, USA, August 20-24,
  2017, Proceedings, Part {I}}, pages 167--187, 2017.

\bibitem{DZ13}
Ivan Damg{\aa}rd and Sarah Zakarias.
\newblock Constant-overhead secure computation of boolean circuits using
  preprocessing.
\newblock In {\em Theory of Cryptography - 10th Theory of Cryptography
  Conference, {TCC} 2013, Tokyo, Japan, March 3-6, 2013. Proceedings}, pages
  621--641, 2013.

\bibitem{DK94}
Iwan~M. Duursma and Ralf K{\"{o}}tter.
\newblock Error-locating pairs for cyclic codes.
\newblock {\em {IEEE} Trans. Information Theory}, 40(4):1108--1121, 1994.

\bibitem{FJN16}
Tore~Kasper Frederiksen, Thomas~P. Jakobsen, Jesper~Buus Nielsen, and Roberto
  Trifiletti.
\newblock On the complexity of additively homomorphic {UC} commitments.
\newblock In {\em Theory of Cryptography - 13th International Conference, {TCC}
  2016-A, Tel Aviv, Israel, January 10-13, 2016, Proceedings, Part {I}}, pages
  542--565, 2016.

\bibitem{GIKW14}
Juan~A. Garay, Yuval Ishai, Ranjit Kumaresan, and Hoeteck Wee.
\newblock On the complexity of {UC} commitments.
\newblock In {\em Advances in Cryptology - {EUROCRYPT} 2014 - 33rd Annual
  International Conference on the Theory and Applications of Cryptographic
  Techniques, Copenhagen, Denmark, May 11-15, 2014. Proceedings}, pages
  677--694, 2014.

\bibitem{Gia16}
Irene Giacomelli.
\newblock {\em New Applications of Secret-Sharing in Cryptography}.
\newblock PhD thesis, Aarhus University, Denmark, October 2016.

\bibitem{Gra}
Markus Grassl.
\newblock Bounds on the minimum distance of linear codes and quantum codes.
  {A}vailable at http://www.codetables.de. {A}ccessed on 01 feb. 2017.

\bibitem{OEIS}
OEIS~Foundation Inc.
\newblock The on-line encyclopedia of integer sequences, https://oeis.org/.

\bibitem{Koe92}
Ralf K\"otter.
\newblock A unified description of an error locating procedure for linear
  codes.
\newblock In {\em Proceedings of Algebraic and Combinatorial Coding Theory,
  Voneshta Voda, Bulgaria}, pages 113--117, 1992.

\bibitem{Lint82}
Jacobus Hendricus~van Lint.
\newblock {\em Introduction to Coding Theory}.
\newblock Springer-Verlag New York, Inc., Secaucus, NJ, USA, 1982.

\bibitem{Mir12}
Diego Mirandola.
\newblock Schur products of linear codes: a study of parameters.
\newblock Master's thesis, Universit\'e de Bordeaux 1 and Stellenbosch
  University, 2012.

\bibitem{MZ15}
Diego Mirandola and Gilles Z{\'{e}}mor.
\newblock Critical {P}airs for the {P}roduct {S}ingleton {B}ound.
\newblock {\em {IEEE} Trans. Information Theory}, 61(9):4928--4937, 2015.

\bibitem{Pel92}
Ruud Pellikaan.
\newblock On decoding by error location and dependent sets of error positions.
\newblock {\em Discrete Mathematics}, 106:369 -- 381, 1992.

\bibitem{Ran13AG}
Hugues Randriambololona.
\newblock Asymptotically good binary linear codes with asymptotically good
  self-intersection spans.
\newblock {\em {IEEE} Trans. Information Theory}, 59(5):3038--3045, 2013.

\bibitem{Ran13S}
Hugues Randriambololona.
\newblock An upper bound of {S}ingleton type for componentwise products of
  linear codes.
\newblock {\em {IEEE} Trans. Information Theory}, 59(12):7936--7939, 2013.

\bibitem{Ran15}
Hugues Randriambololona.
\newblock On products and powers of linear codes under componentwise
  multiplication.
\newblock {\em Contemp. Math.}, 637:3--78, 2015.

\bibitem{Min}
Rudolf Sch{\"u}rer and Wolfgang~Ch. Schmid.
\newblock {\em MinT: A Database for Optimal Net Parameters.}
\newblock In Harald Niederreiter, Denis Talay (eds) Monte Carlo and Quasi-Monte
  Carlo Methods 2004. Springer, Berlin, Heidelberg. Database available at
  http://mint.sbg.ac.at/. {A}ccessed on 01 Feb. 2017.

\bibitem{STV92}
Igor~E. Shparlinski, Michael~A. Tsfasman, and Serge~G. Vladut.
\newblock Curves with many points and multiplication in finite fields.
  {P}roceedings of the {I}nternational {W}orkshop held in {L}uminy, {F}rance,
  {J}une 17–21, 1991.
\newblock In {\em Coding Theory and Algebraic Geometry.}, pages 145--169, 1992.



\end{thebibliography}
\end{document}